\documentclass[12pt]{iopart}
\usepackage{iopams}  

\usepackage{graphicx}
\usepackage{color}
\usepackage{bm}
\usepackage{appendix}

\newcommand{\ve}{\varepsilon}
\newcommand{\olap}[2]{\left<#1\right.\left|#2\right>}

\begin{document}

\title[Microscopic bosonization of band structures]{Microscopic bosonization of
band structures:\\ X-ray processes beyond the Fermi edge}

\author{Izak Snyman}
\address{Mandelstam Institute for Theoretical Physics, School of Physics, University 
of the Witwatersrand, Wits, 2050, South Africa}
\author{Serge Florens}
\address{Institut N\'{e}el, CNRS and Universit\'e Grenoble Alpes, F-38042 Grenoble, France}

\begin{abstract} 
Bosonization provides a powerful analytical framework to deal with one-dimensional strongly 
interacting fermion systems, which makes it a cornerstone in quantum many-body theory. 
Yet, this success comes at the expense of using effective infrared parameters, and 
restricting the description to low energy states near the Fermi level. We propose a radical 
extension of the bosonization technique that overcomes both limitations, allowing computations 
with microscopic lattice Hamiltonians, from the Fermi level down to the bottom of the band.
The formalism rests on the simple idea of representating the fermion kinetic term in the 
energy domain, after which it can be expressed in terms of free bosonic degrees of freedom. 
As a result, one- and two-body fermionic scattering processes generate anharmonic 
boson-boson interactions, even in the forward channel. We show that up to moderate interaction strengths,
these non-linearities can be treated analytically at all energy scales, using 
the x-ray emission problem as a showcase. In the strong interaction regime, we employ a 
systematic variational solution of the bosonic theory, and obtain results that agree quantitatively 
with an exact diagonalization of the original one-particle fermionic model. This provide 
a proof of the fully microscopic character of bosonization on all energy scales for an 
arbitrary band structure. Besides recovering the known x-ray edge singularity at the 
emission threshold, we find strong signatures of correlations even at emmision frequencies beyond 
the band bottom. 
\end{abstract}

\maketitle
\section{Introduction}
\label{sec1}

The general description of fermions at finite density constitutes a central problem
in physics, requiring a microscopic understanding of a
macroscopically large number of interacting particles. Standard ways to simplify
the task often rely on a description of the low energy excitations
above the many-body ground state in terms of weakly interacting quasi-particles
with effective parameters. Predicting the behavior of excitations beyond this
infrared regime remains challenging, and
numerical techniques are typically used, with various limitations regarding
the number of handled particles, the range of accessible temperatures, the strength 
of the Coulomb interaction, and the resolution of fine spectroscopic structures~\cite{Sandvik,Schollwoeck,Bulla,White}.
However, modelling the response of strongly correlated solids up to large excitation
energies has become a pressing issue, especially with advances in spectroscopic 
methods such as inelastic neutron scattering~\cite{neutrons}, angle-resolved
photoemission~\cite{ARPES}, and resonant inelastic x-ray
scattering~\cite{RIXS}. In the field of cold atoms, various techniques
such as Bragg~\cite{Bragg} and momentum-resolved Raman~\cite{Raman} spectroscopies
have also been developed to probe the spectrum of interacting Fermi gases.

At finite densities and low energies, one-dimensional fermion systems are
elegantly described in terms of density fluctuations that behave like bosonic
particles \cite{Giam,Gogolin,Haldane}. The mapping between fermions and bosons,
known as bosonization, is usually applied in conjunction with a linearization of
the fermion dispersion relation, which produces a low energy effective theory in
which wavelengths of the order of the lattice constant are integrated out. Since
information about the crystal lattice structure is lost in the process, it is
difficult to describe phenomena resulting from an interplay between band
stucture and many-body correlations. Incorporating band structure in the
conventional bosonic picture introduces boson-boson interactions with a
divergent perturbative
expansion~\cite{Samokhin,Teber,Aristov,Imambekov,Matveev}. This has cast serious
doubt on the usefulness of the bosonic picture beyond the infrared limit. In
this paper, we take a new approach to bosonization by mapping free fermions with
an arbitrary spectrum onto well-crafted free bosonic degrees of freedom. Unlike
conventional bosonization, our approach is readily applicable to microscopic
rather than low-energy effective models, either on tight-binding lattices 
or for continuous wave equations with a non-linear dispersion.

To demonstrate the utility of our approach, we investigate the stimulated
emission of x-rays when an electron in a metal decays into a core-orbital inside
one of the lattice ions, the so-called x-ray edge problem~\cite{Ohtaka,Rehr} .
For low frequency emitted photons (close to the Fermi edge), conventional
bosonization provides an elegant answer~\cite{Schotte}. There are however
interesting effects, such as multiple-electron processes that produce a non-zero
emission rate at frequencies beyond the {\em band} edge, that conventional
bosonization cannot hope to capture. Using our new approach to bosoniztion, we
calculate the full emission spectrum, from the low energy threshold (the edge)
up to the high energy regime where band structure plays an important role. For
weak interactions, we obtain for the first time an analytical solution that 
incorporates band structure effects. Beyond this regime, we implement a
non-perturbative variational approach that yields near-perfect agreement with
brute force numerical diagonalization (the current state of the art). Besides
shedding light on the nature of many-body correlations and the natural degrees
of freedom in the problem, our approach is more efficient than existing ones.
The scaling of computation time with system size $\Omega$ is $\Omega\ln \Omega$
as opposed the the significantly more expensive $\Omega^3$ of the brute force
approach~\cite{Markiewicz,Knap}.

While we restrict our analysis to the specific physical problem of
x-ray emission, in order to showcase our method, the ideas developed are general. 
A large class of low-dimensional fermion systems can be mapped onto new bosonic 
problems that faithfully reproduce physics from the infrared to the ultraviolet. 
While generically the bosonic models are interacting, they hold great advantages 
over the original fermionic description, because here the interactions do not invalidate
a quasiparticle description in terms of bosons. In this work, we have developed
a systematic variational treatment of the bosonized theory, since the x-ray
response can be obtained solely from the knowledge of the underlying
wavefunction. We anticipate that microscopic bosonization provides
bosonic models that may be more efficiently treated by standard 
numerical methods than their original fermionic versions, since the interplay of band-structure and
x-ray thresholds or Tomonaga-Luttinger physics
is already incorporated at 
quadratic order in the bosonic theory.

The paper is organized as follows. 
Section~\ref{sec2N} develops a general microscopic bosonization approach to cope 
with arbitrary electronic band structures, with a detailed reformulation of
standards electronic operators in bosonic form.
Section~\ref{sec3N} discusses general aspects of the x-ray problem, presenting the 
microscopic Hamiltonian, the relevant optical response function, and the low-energy physics 
resulting from the orthogonality catastrophe. 
In Section~\ref{sec4N}, we use the microscopic bosonization method to derive an analytical solution 
of the x-ray edge problem for an arbitrary band structure, in the case where the
interaction strength is not too strong. We show how conventional
bosonization results are recovered for the threshold singularities in the case of a linear
dispersion. 
In Section~\ref{sec5N}, the case of large interaction is then addressed by more advanced many-body wave 
function methods. This involves natural extensions of the analytical
theory based on variational coherent states of the normal bosonic modes. 
In closing we provide a perspective that underlines the many possibilities that 
microscopic bosonization could open in the field of strongly correlated systems.

\section{Microscopic bosonization formalism for lattice models}
\label{sec2N}
\subsection{Arbitrary electronic spectrum in linear form}

Our starting point is the free Hamiltonian for a single species of fermion
in a one-dimensional crystal
\begin{equation} 
H_0 = \int_{0}^{\pi} dq \; \ve(q) \tilde c^\dagger_q \tilde c_q,
\end{equation} 
with $\tilde c^\dagger_q$ the fermion creation operator for the
even band orbital that is a linear combination of Bloch states with crystal 
momenta $q$ and $-q$. 
We choose to work here with even modes (discarding odd
modes) anticipating the fact that we will later consider a single-site 
impurity in a time-reversal invariant crystal, that scatters only even states. 
Alternatively, it is also possible to work with chiral species $\tilde c^\dagger_{q>0}$ 
and $\tilde c^\dagger_{q<0}$, if the problem at hand requires.
Operators associated with the momentum and position representations 
are denoted with tildes, in order to distinguish them from the operators associated 
with the energy representation (and its conjugate time representation defined below).
We will assume that the dispersion relation $\ve(q)$ increases monotonically 
for $q\in(0,\pi)$, but that its precise form is arbitrary. For a non-monotonous
dispersion relation, several electronic sub-bands would have to be introduced.

Instead of linearizing the dispersion relation around the Fermi energy, as one normally
does when bosonizing, we invert it to express the momentum $q$ as 
a function of energy $\ve$. We take the bottom and top of the band to lie at $-D$ and $D$ respectively. 
For $\ve\in(-D,D)$, we thus define new fermionic operators 
\begin{equation}
c_{\ve}^\dagger=\sqrt{2\pi N(\ve)}\tilde c_{q(\ve)}^\dagger,
\label{credef}
\end{equation}
where 
$N(\ve)=\frac{1}{2\pi} \frac{dq(\ve)}{d\ve}$
is the density of even states per unit length, 
so that their anticommutator reads $\{c_\ve,c_{\ve'}^\dagger\}=\delta(\ve-\ve')$ and
\begin{equation}
H_0=\int_0^\pi \!\!dq\, \varepsilon(q) :\tilde c_q^\dagger \tilde c_q: =
\int_{-D}^D \!\!d\ve\, \ve :c_{\ve}^\dagger c_{\ve}:.
\end{equation}
This is formally similar to the operator $\int_{-\infty}^\infty dq\,q\,:\tilde 
c_q^\dagger \tilde c_q:$ that is encoutered in the case of the linear
dispersion~\cite{Giam}, but with one important difference: The band of the 
$c_{\ve}^\dagger$ fermions is bounded. However, one can always view 
a given system as a sub-system that is uncoupled from the rest of a larger system.
In order to extend the energy spectrum infinitely downwards and upwards, we
introduce spectator fermions created by operators $c_\ve^\dagger$, with $|\ve|> D$, that 
anti-commute with the physical fermion operators within the band, and whose Hamiltonian reads 
$H_{\rm spec}=\int_{|\ve|>D} \!d\ve\; \ve : c_\ve^\dagger c_\ve:$. Here normal ordering is 
with respect to the state in which orbitals with $\ve<-D$ are occupied and ones with 
$\ve>D$ are empty. Rescaling transformations such as $\tilde c_q\to c_\ve$ is standard in 
the analysis of infinite systems~\cite{Bulla}, but enlarging the Hilbert space 
with spectator orbitals is a new ingredient that is crucial for the microscopic
bosonization approach.
The Hamiltonian for the enlarged system reads
\begin{eqnarray}
H_{0,{\rm enl}}&=H_0+H_{\rm spec}=
\int_{-\infty}^\infty \!\!d\ve \; \ve :c_{\ve}^\dagger c_{\ve}: .
\label{h_0enl}
\end{eqnarray}
Due to the normal ordering of the spectator fermion operators, $H_{0,{\rm enl}}$ 
has the same ground state energy as $H_0$.
In the above expression (\ref{h_0enl}), it is manifest that $H_{0,{\rm enl}}$
does not couple the spectator fermions and the band fermions. Thus, in the physical 
sector, the enlarged free Hamiltonian is fully equivalent to the original fermionic model.

\subsection{Free bosonic representation of the electronic band}
The second important step is to obtain a representation of the 
free electronic Hamiltonian $H_{0,{\rm enl}}$ (with an arbitrary non-linear dispersion) 
in terms of free bosonic degrees of freedom. 
With the energy representation~(\ref{h_0enl}) as our starting point, we define conjugate 
time representation operators through the Fourier transform:
\begin{equation}
\psi(\tau)=\int_{-\infty}^\infty \!\!\frac{d\ve}{\sqrt{2\pi}}\, e^{i\ve\tau} c_\ve,
\label{Fourier}
\end{equation}
so that $\{\psi(\tau),\psi^\dagger(\tau')\}=\delta(\tau-\tau')$. Thus, in the time
representation, the enlarged electronic band Hamiltonian~(\ref{h_0enl}) reads
\begin{equation}
\label{H0timedomain}
 H_{0,{\rm enl}} = -i \int_{-\infty}^{+\infty} \!\!d\tau \, \psi^\dagger(\tau)
\partial_\tau \psi(\tau).
\label{h0tau}
\end{equation}
The presence of the first order differential operator $-i\partial_\tau$ in
Eq.~(\ref{H0timedomain}), reminiscent of a linearly dispersing wave equation,
paves the way for expressing the exact kinetic energy
of band electrons as a single-particle additive bosonic operator.
In analogy with the standard bosonization identities~\cite{Giam} for a linearized
spectrum in momentum representation, we construct bosonic annihilation operators
\begin{equation}
b_\ve=\frac{1}{\sqrt{\ve}}\int_{-\infty}^\infty \!\!d\tau\, 
e^{-i\ve\tau}\psi^\dagger(\tau)\psi(\tau),~~~\ve>0,\label{bdef}
\end{equation}
that, in our case, are associated with the energy representation. 
This definition constitutes the essence of the microscopic bosonization method.
In terms of these bosonic operators, the free electronic kinetic 
term~(\ref{h0tau}) reads simply 
\begin{equation}
H_{0,{\rm enl}} = \int_0^\infty \!\!d\ve\;
\ve\, b_\ve^\dagger b_\ve, 
\label{H0boso}
\end{equation}
similarly to the case of the linearized theory. 

\subsection{Construction of the lattice fermionic operators}
The final step of the microscopic bosonization is to express the local
electronic fields of the original lattice in terms of the bosons associated with the energy 
representation~(\ref{bdef}). Without loss of generality, we consider the
electronic field for the site at the origin:
\begin{equation}
\tilde \psi_0^\dagger = \int_0^\pi \frac{dq}{\sqrt{\pi}} \; \tilde c_q^\dagger 
= \int_{-D}^D d\ve\, \sqrt{2N(\ve)} c^\dagger_\ve,
\label{localfield}
\end{equation}
where we have used Eq.~(\ref{credef}) to go from the momentum to the energy
representation. 
From the Fourier transform~(\ref{Fourier}), we 
can then obtain a faithful expression of the lattice field in terms of the 
time-local field:
\begin{equation}
\hspace{-1cm}
\tilde \psi_0^\dagger 
= \int_{-D}^D d\ve\,\sqrt{2N(\ve)}
\int_{-\infty}^{+\infty} \frac{d\tau} {\sqrt{2\pi}}\;
e^{i\ve\tau} \psi^\dagger(\tau)
=2\sqrt{\pi} \int_{-\infty}^{+\infty} d\tau\; \Delta(\tau) \psi^\dagger(\tau),
\label{psiinter}
\end{equation}
where we have defined 
\begin{equation}
\Delta(\tau)=\int_{-D}^{+D} \frac{d\varepsilon}{2\pi}\,
e^{i\varepsilon\tau}\sqrt{N(\varepsilon)}.\label{ddef}
\end{equation}
Note that for a strictly infinite linear dispersion, the function $\Delta(\tau)$ 
becomes nearly local as the density of states $N(\ve)$ is constant on all
energy scales (up to the inverse of the short time cut-off given by the bosonization 
regulator $a$). 

Finally, we can express the fermionic operators in the conjugate 
time representation using the usual bosonization identity~\cite{Giam}
\begin{equation}
\psi(\tau)=\frac{U}{\sqrt{2\pi a}}e^{i\ve_F\tau+\int_0^\infty\!\!\frac{d\ve}{\sqrt{\ve}}\,
e^{-a\ve/2}\left(b_\ve e^{i\ve\tau}-b_\ve^\dagger e^{-i\ve\tau}\right)},
\label{bosodef}
\end{equation}
in terms of the bosonic modes in the energy domain. In this expression, $U$ 
is the usual unitary Klein factor, and $\ve_F$ is the Fermi energy. It is crucial 
to note that the ultraviolet cut-off $1/a$ is assumed here to be much larger than 
the band width $2D$. This is in sharp contrast to the case of the linearized system, 
were $1/a$ represents the boundary between kept and discarded modes, which must lie 
sufficiently deep inside the band, for the linear approximation to the dispersion 
relation to hold. In the expressions that appear below, we will take the limit 
$a\to 0^+$, which here is well-behaved, because now the half-bandwidth $D$ plays the 
role of a natural physical cutoff. From Eqs.~(\ref{psiinter}-\ref{bosodef}) we obtain 
a microscopic expression of the lattice fermion creation operator in terms of the
collective bosonic fields:
\begin{equation}
\tilde \psi_0^\dagger = \sqrt{\frac{2}{a}}\,U^\dagger \int_{-\infty}^{+\infty} d\tau\; \Delta(\tau)^*
e^{-i\ve_F\tau-\int_0^\infty\!\!\frac{d\ve}{\sqrt{\ve}}\,
e^{-a\ve/2}\left(b_\ve e^{i\ve\tau}-b_\ve^\dagger e^{-i\ve\tau}\right)}.
\label{psi0final}
\end{equation}

It is also useful to provide an explicit expression for the local electronic
density operator on the lattice:
\begin{equation}
\hspace{-2cm}
\tilde \psi_0^\dagger \tilde \psi_0 = \frac{1}{2}+2
\int_{-\infty}^{\infty}\!\!d\tau_1 
\int_{-\infty}^{\infty}\!\!d\tau_2\,\mathcal P
\frac{e^{i\ve_F(\tau_2-\tau_1)}}{i(\tau_2-\tau_1)}
\Delta(\tau_1)\Delta(\tau_2)^*
e^{i\left[\varphi^\dagger(\tau_1)-\varphi^\dagger(\tau_2)\right]}
e^{i\left[\varphi(\tau_1)-\varphi(\tau_2)\right]},\label{rho_boson}
\end{equation}
in terms of the bosonic operators in the time domain:
\begin{equation}
\varphi(\tau)=-i\int_0^\infty \!\!\frac{d\varepsilon}{\sqrt{\varepsilon}}\,
e^{i\varepsilon \tau}b_\varepsilon.
\end{equation}
Note that in Eq.~(\ref{rho_boson}), boson operators are normal-ordered and 
limit $a\to0^+$ has been taken.
Eqs.~(\ref{H0boso}), (\ref{psi0final}) and (\ref{rho_boson}) are the central result of 
this section, allowing us to express faithfully standard lattice fermion operators
in terms of collective boson modes that leave the kinetic energy as a purely harmonic term.
The approach that we developed here is thus completely general, and could be applied to 
bosonize arbitrary one-dimensional models (including bulk interaction among fermions). 
We now turn in the next section to the simplest testbed for these ideas, namely the 
x-ray edge singularity in metals. This will allow us to extend previous analytical
techniques~\cite{Schotte} for an arbitrary density of states, and demonstrate
the microscopic equivalence on all energy scales of the lattice fermionic model to 
our bosonic representation.

\section{Formulation and review of the x-ray edge problem}
\label{sec3N}

\subsection{Motivation}
The x-ray edge problem concerns the stimulated transition of a fermion between a Fermi sea and a
localized orbital ~\cite{Ohtaka,Rehr}. In the solid state context, the
electromagnetic radiation that stimulates the transition consists of soft
x-rays, and we will use this terminology, regardless of the actual wavelength.
Due to initial or final state interactions between the Fermi sea and the
localized orbital, the x-ray transition rates involve the overlap between the
calm Fermi sea, and one that is agitated. Such overlaps, and the associated
Anderson orthogonality catastrophe~\cite{PWAcatastrophe}, lead to a power-law
singularity in x-ray emission and absorption spectra,
close to the Fermi threshold.
The low-energy physics of the Fermi edge singularity was elucidated theoretically in the
Sixties~\cite{Mahan,Nozieres1,Nozieres2,Nozieres3} by Nozi\`eres and others. Full
spectroscopic calculations can be performed nowadays using brute force numerical 
diagonalization on large systems~\cite{Markiewicz,Knap}, or using approximate
diagrammatic methods~\cite{Delft}.
One physical question that we wish to address here
concerns the spectroscopic signatures of many-body physics away from the
Fermi level, which cannot be described by Nozi\`eres's low energy theory.
Besides their obvious relevance in realistic aspects of x-ray spectra~\cite{Liang}, 
it is worth mentioning that initial or final state interaction effects due to dynamical impurities can also 
occur in mesoscopic devices~\cite{Hentschel} and in cold atom gases~\cite{Knap,Cetina}.

Historically, the x-ray problem represents an early success for the application 
of bosonization. The bosonic description~\cite{Schotte} proves more succinct 
than the fermionic one~\cite{Nozieres3}, provided one accepts 
as starting point an effective low energy model with renormalized parameters,
which are often difficult to express in terms of the original microscopic ones.
In contrast, the parameters that appear in our bosonization approach, are the original
microscopic ones.
In principle, our microscopic bosonization approach could be used in conjunction with a variety 
of existing numerical techniques to compute density matrices, partition functions, or arbitrary 
dynamical response functions.
Here we take a non-perturbative 
variational approach that is also physically intuitive. 
Since coherent states constitute a natural language for orthogonality catastrophe physics,
we propose variational states based on multi-mode bosonic coherent
states, which provide excellent x-ray emission spectra on all energy scales, and an appealing 
description of many-fermion states in the presence of dynamic impurities.

\subsection{Modelling of a Fermi sea with a dynamical impurity}

We start with a description of our model, which consists of a single band of a
one-dimensional crystal and a nearby localized orbital. (In the atomic gas
context, the crystal would be engineered using an optical lattice.) The band is
partially filled with non-interacting fermionic particles. The localized
orbital can either be empty or filled. In the absence of x-ray stimulation
(which we discuss in the next subsection), it is assumed that particles do not
tunnel between the band and the localized orbital. When the localized orbital
is empty, fermions in the crystal undergo potential scattering, and when the
localized orbital is filled, the fermions in the crystal are not scattered.
This form of interaction naturally arises in the solid state context if the
localized orbital represents a state in a core shell of one of the lattice ions.
In this case, the empty orbital state corresponds
to a core hole that produces an attractive Coulomb potential.
From a mathematical point of view however,
the roles of the empty and filled states of the localized orbital can be
reversed without affecting the applicability of our method. Since spin plays no
role, we consider a single spin species.
We assume that the static potential of
the empty localized orbital is localized to site zero of the crystal.
We assume inversion symmetry, so that spatial parity is a good quantum number, and only even parity band-orbitals
couple to the localized orbital. We remove odd parity band-orbitals from our
description at the outset. We define normal ordering $:\ldots :$ with respect
to the clean Fermi sea associated with the band when it does not interact with
the localized orbital. We measure crystal momentum $q$ in inverse units of the lattice constant.
The system is thus described by the unperturbed Hamiltonian
\begin{equation}
H=\left(H_V+\frac{V}{2}-\ve_d\right)dd^\dagger+H_0 d^\dagger d,
\end{equation}
where $H_0=H_{V=0}$ with
\begin{equation}
H_V =
\int_0^\pi \!\!dq\, \varepsilon(q) :\tilde c_q^\dagger \tilde c_q:
+V\left(\tilde \psi_0^\dagger \tilde \psi_0-\frac{1}{2}\right),\label{fermion_H}
\end{equation}
and
\begin{equation}
\tilde \psi_0=\frac{1}{\sqrt{\pi}}\int_0^\pi \!\!dq\, \tilde c_q. \label{def_psi0}
\end{equation}
The operator $d^\dagger$ creates a fermion in a localized orbital with energy
$\varepsilon_d$. The operator $\tilde c_q^\dagger$ creates a fermion with
positive momentum $q$ in the even band orbital, as defined in the previous section.
The operator $\tilde \psi_0^\dagger$, also defined in the previous section, creates a fermion in the Wannier
orbital associated with the site zero of the lattice. Again, the form of the
electronic density of state is arbitrary. In the soft x-ray emission and absorption 
problem in metals, the orbital at energy $\varepsilon_d$ describes a core level, so 
that it lies below the bottom of the band in energy, and the interaction is attractive, 
{\it i.e.} $V<0$. We are however interested in the model for its own sake, and will not 
place any restrictions on $\varepsilon_d$ or $V$. The constant term $-V/2$ in the 
definition of $H_V$ is included for later convenience.

\subsection{X-ray transition rate}

We consider an initial state in which the localized orbital interacts with the particles in the crystal
and the particles in the crystal have reached zero temperature equilibrium.
The system is then subjected to incoherent electromagnetic radiation at frequency $|\omega|$, 
that stimulates a transition in which a fermion tunnels
between the band and the localized orbital.
We assume that the radiation only stimulates tunnelling between the localized orbital and
the Wannier orbital associated with the site zero of the lattice, {\em i.e.} we neglect 
a possible momentum dependence of the optical matrix elements. 

According to Fermi's golden rule, the transition rate for this process is given by
\begin{equation}
W=\gamma^2\sum_{\nu} \left|\left<\Psi^0_\nu\right|\tilde \psi_0\left|\Psi_0^V\right>\right|^2
\delta(E_\nu^0+\varepsilon_d+\omega -E_0^V),\label{rate}
\end{equation}
where $\gamma$ is a tunnelling amplitude (with dimensions of energy). Here,
$\left|\Psi_0^V\right>$ is the ground state (Fermi sea) of $H_V$, and $E_0^V$
the associated energy, in the sector of Fock space that contains the same number
of particles as when all single particle orbitals of $H_0$ up to the Fermi
energy are filled. The states $\left|\Psi_\nu^0\right>$ are the complete set of
eigenstates of $H_0$, in the sector of Fock space with one less particle than
$\left|\Psi_0^V\right>$, and $E_\nu^0$ are their energies. When $\omega>0$ in
(\ref{rate}), tunnelling is accompanied by the stimulated emission of a photon of
frequency $\omega$, and when $\omega<0$, by the absorption of a photon of energy
$-\omega$. We define a shifted frequency 
\begin{equation}
\varepsilon=\omega+\varepsilon_d-E_0^V,
\end{equation}
and consider the transition rate $W$ as a function of $\varepsilon$. Note that,
due to normal ordering, the ground state energy of $H_0$ is zero. Thus, the
transition rate $W(\varepsilon)$ vanishes for $\varepsilon>0$. At the threshold
$\varepsilon=0$, the final state of the crystal is the clean Fermi sea.

By writing the $\delta$-function in (\ref{rate}) as 
\begin{equation}
\delta(\varepsilon+E_\nu^0)=\frac{1}{\pi}{\rm Re}\int_0^\infty \!\!dt\, e^{-i(\varepsilon+E_\nu^0)t},
\end{equation}
and using the completeness of the states $\left|\Psi_\nu^0\right>$, we can rewrite formula 
(\ref{rate}) for the transition rate as
\begin{eqnarray}
W(\varepsilon)&=&\frac{\gamma^2}{\pi}{\rm Re}\int_0^\infty \!\!dt\, e^{-i\varepsilon t} P(t),\\
P(t)&=&\left<\Psi_0^V\right|\tilde \psi_0^\dagger e^{-i H_0 t}\tilde \psi_0\left|\Psi_0^V\right>.\label{pt}
\end{eqnarray}

For $V=0$, the transition rate evaluates to
\begin{equation}
W(\ve)=2\gamma^2 N(\ve_F+\ve)\theta(-\ve),\label{v0res}
\end{equation}
where 
\begin{equation}
N(\ve)=\frac{1}{2\pi} \frac{dq(\ve)}{d\ve}
\end{equation}
is the density of even states per unit length, and $\ve_F$ is the Fermi energy.
At the threshold value of $\ve=0$, the transition rate makes a discontinuous
jump. Just below the threshold, the transition rate is finite. For the
one-dimensional band that we consider, the transition rate diverges at $\ve$
equal to the energy difference between the bottom of the band and the Fermi
energy. This is due to the van Hove singularity in $N(\ve)$ at the bottom of
the band. For lower energies, the transition rate is zero (to order $\gamma^2$).

For non-zero $V$, the transition rate develops a prominent feature close to the
$\ve=0$ threshold. At energies $\ve<0$ such that $|\ve|$ is much less than both
the energy differences between the top edge of the band and the Fermi energy,
and between the Fermi energy and the bottom edge of the band, the transition
rate acquires a power-law form (the so-called Fermi-edge
singularity)~\cite{Mahan, Nozieres1, Ohtaka}
\begin{equation}
W(\ve)\propto \left|\frac{\ve}{\ve_F}\right|^{[(f_0+1)^2-1]}\theta(-\ve).\label{powerlaw1}
\end{equation}
Here $ f_0=\phi(\ve_F)/\pi$ with
\begin{equation}
\phi(\ve)=\arctan\left[\frac{2\pi N(\ve) V}{1+2V\mathcal P
\int_{-\infty}^\infty \!\!d\ve'\,\frac{N(\ve')}{\ve'-\ve}}\right]\label{phaseshift}
\end{equation}
the exact scattering phase shift that the local potential
$V\tilde\psi_0^\dagger\tilde\psi_0$ induces on a fermion in the band incident at
energy $\ve$, so that, according the the Friedel sum rule, $f_0$ is the average
number of particles displaced by the potential. For negative $\phi$, the
transition rate close to threshold is enhanced, while for positive $\phi$, it is
a suppressed. 

Further below the $\ve=0$ threshold, the transition rate is also
modified in important ways, but these effects are not accounted for by
Nozi\`eres's approach. For instance, transitions to final states containing
multiple particle-hole excitations above the clean Fermi sea will cause a
non-zero transition rate below the bottom of the band. In order
to exactly calculate $W(\ve)$ at values of $\ve$ that are finite compared to the
bandwidth, one has to employ brute force numerics~\cite{Knap}. Each evaluation of
$P(t)$ in Eq.~(\ref{pt}) involves the numerical evaluation of a Slater determinant, 
and other matrix operations, for which the computation time scales as the third power 
of the number of particles in the system. We provide more detail about this approach 
in the supplementary material that accompanies this Article. 

Apart from the transition that we consider, namely one in which the initial
state of the localized orbital interacts with the impurity and the final one
does not, one can also consider the case where the interaction occurs in the
final state. In this case, the initial state of the band is the clean Fermi sea
$\left|\Psi_0^0\right>$, and after a particle is added to site zero, the system
evolves with $H_V$. The wave function method we develop below is tailored to
time evolution with $H_0$ rather than $H_V$, and we therefore consider here
only the initial and not the final state interaction situation.

\section{Weak interaction solution of the x-ray edge problem on all energy scales} 
\label{sec4N}

\subsection{Bosonized form of the x-ray edge model with arbitrary density of states}

We express the electronic x-ray edge Hamiltonian~(\ref{fermion_H})
 in terms of collective boson degrees of freedom, using the 
equations~(\ref{H0boso}) and (\ref{rho_boson}). We readily obtain 
\begin{eqnarray}
\hspace{-2.5cm}
\nonumber
H_{V,{\rm enl}}=\int_0^\infty \!\!\!\!\!\!d\ve\,\ve b_\ve^\dagger b_\ve 
+ 2V \int_{-\infty}^{\infty}\!\!\!\!\!\!d\tau_1 
\int_{-\infty}^{\infty}\!\!\!\!\!\!d\tau_2\,
\mathcal P
\frac{e^{i\ve_F(\tau_2-\tau_1)}}{i(\tau_2-\tau_1)}
\Delta(\tau_1)\Delta(\tau_2)^*
e^{i\left[\varphi^\dagger(\tau_1)-\varphi^\dagger(\tau_2)\right]}
e^{i\left[\varphi(\tau_1)-\varphi(\tau_2)\right]}.\!\!.\\
\label{h_boson}
\end{eqnarray}
We stress that the bosonic form of $H_{V,{\rm enl}}$ does not couple band fermions
and spectator fermions, although this is no longer manifest, as these states are
mixed in a complicated way within the bosonic fields. In its bosonized form,
Hamiltonian~(\ref{h_boson}) can in principle be analyzed in a variety
of ways. For instance, numerical renormalization group~\cite{Freyn} or quantum Monte 
Carlo~\cite{Moon,Leung,Hamamoto,Rosch} calculations have previously been developed to deal 
with similar Hamiltonians. 
We choose to focus in what follows on wave function-based methods, because according to Eq.~(\ref{pt}), 
we only need to find the ground state of $H_{V,{\rm enl}}$ in order to calculate 
the transition rate $W(\ve)$. Since $H_{V,{\rm enl}}$ is not quadratic in the bosonic 
degrees of freedom, there is probably no simple expression for the exact ground state 
in the bosonic language. Our main task will be in what follows to provide controlled 
analytical and numerical computation of this bosonic many-body state.

\subsection{Analytical solution at moderate interaction for an arbitrary band}
\label{ansol}
We provide here a controlled analytical solution of the x-ray problem in an 
arbitrary band and on all energy scales (even beyond the band
edge), provided that the impurity interaction is weak enough. The orthogonality 
catastrophe induced by the impurity will be treated non-perturbatively, thanks 
to the bosonic representation~(\ref{h_boson}) of the problem.
This solution will also serve as a starting point for further variational
refinements that we will develop in the following section in order to address
the regime of strong coupling.
The starting point is the observation that for a weak interaction strength 
$V\ll D$, the bosonic modes $b_\ve^\dagger$ in~(\ref{h_boson}) fluctuate 
very mildly around their undisplaced configuration $b_\ve^\dagger=0$. Thus, 
it is legitimate to expand the exponential term in~(\ref{h_boson}) to first order
in the bosonic modes. This produces a quadratic bosonic Hamiltonian, reminiscent of the
case with linear dispersion~\cite{Schotte,Giam}, while still encoding non-trivial 
features of the full band structure, in contrast to the usual bosonization solution.
In this approximation, the exact Hamiltonian matrix element for creating a single 
particle-hole pair of energy $\ve$ is included in the description, for all
energies $\ve$ up the extreme ultraviolet limit $2D$.
The x-ray problem is thus accurately captured at an arbitrary energy, as long as
$|V|$ is sufficiently small. 

In order to visualize the configuration of the bosonic degrees of freedom it is useful 
to define a rescaled average bosonic displacement
\begin{equation}
\Phi(\ve)=-\frac{\sqrt{\ve}}{2}\left<b_\ve+b_\ve^\dagger\right>\label{opar}
\end{equation}
with respect to the exact bosonic ground state, a quantity which vanishes 
when the potential $V$ is turned off. In the case of a linear dispersion relation, 
$\Phi(\ve=v_F k)$ corresponds to the $k$-component of the Fourier transform of 
the fermion density ($v_F$ is the Fermi velocity). In general,
$\Phi(0)$ equals the charge displaced by the impurity potential, which according
to the Friedel sum rule, equals the phase shift at the Fermi energy, divided by
$\pi$. The bosonic displacement can be calculated by exact diagonalization, in the
original fermionic representation of $H_V$, by using the definition (\ref{bdef})
for $b_\ve$. This reveals that its maximum value as a function of $\ve$ is
always of the order of $\Phi(0)$. Thus, $\Phi(\ve)$ is small when the impurity
induces a small phase shift at the Fermi energy, namely when $V$ is small
enough. This confirms our initial argument that expanding the exponentials
$e^{-i\left[\varphi^\dagger(\tau_1)-\varphi^\dagger(\tau_2)\right]}
e^{-i\left[\varphi(\tau_1)-\varphi(\tau_2)\right]}$ in (\ref{h_boson}) to first
order in boson operators will yield an accurate approximation. The
approximation should work particularly well for a particle-hole symmetric band
at half-filling, where the dropped terms in the expansion contain at least three
normal-ordered boson operators. Whereas the linear dispersion
approximation~\cite{Schotte,Giam} can only predict $\Phi(0)$ and the power-law 
behavior of the transition rate $W(\ve)$ close to the Fermi edge threshold, here 
we expect to obtain quantitatively correct results for $\Phi(\ve)$ and $W(\ve)$ 
for all $\ve$ as long as the potential strength $V$ is sufficiently small.

Expanding the potential term in (\ref{h_boson}) to first order in $b_\ve$ and 
$b_\ve^\dagger$, we obtain the lowest order Hamiltonian
\begin{equation}
\hspace{-1.5cm}
H_{V,{\rm enl}}\simeq \int_0^\infty \!\!d\ve \, \ve\left[b^\dagger
+\frac{f_1(\ve)}{\sqrt{\ve}}\right]\left[b^\dagger+\frac{f_1(\ve)}{\sqrt{\ve}}\right]
+V\left(\left<n\right>-\frac{1}{2}\right)
-\int_0^\infty \!\!d\ve [f_1(\ve)]^2,
\end{equation}
where $\left<n\right>=2\int_{-D}^{\ve_F}d\ve\,N(\ve)$ is the number of particles per lattice 
cell of the clean band, and the displacement
\begin{equation}
\hspace{-1.5cm}
f_1(\varepsilon)=\frac{2V}{\varepsilon}\int_0^\infty \!\!d\omega\,\sqrt{N(\ve_F-\omega)}
\left[\sqrt{N(\ve_F-\omega+\ve)}-\sqrt{N(\ve_F-\omega-\ve)}\right].\label{pert}
\end{equation}
In this approximation, the normalized ground state is a coherent state
\begin{equation}
\left|f_1\right>=\exp\left\{-\int_0^\infty \!\!\frac{d\varepsilon}{\sqrt{\varepsilon}}\,
f_1(\varepsilon)\left[b_\varepsilon^\dagger- b_\varepsilon\right]\right\}
\left|{\rm vac}\right>,\label{coherent}
\end{equation}
parametrized by the displacement~(\ref{pert}).
While this wavefunction is similar in form to that of the linearly dispersive
model~\cite{Schotte,Giam}, the average oscillator displacements $f_1(\ve)$ now
clearly encode information about the whole band structure.
In particular, from (\ref{pert}) follows that $f_1(\ve)$ vanishes for $\ve>2D$
and that, owing to the van Hove singularities in $N(\ve)$ at $\ve=\pm D$,
$f_1(\ve)$ has cusps at $\ve=D\pm\ve_F$. As we will see below, these are properties 
shared with the exact solution of the problem. 

Within this first order approximation, the rescaled bosonic displacement~(\ref{opar}) 
is simply given by $\Phi(\ve)\simeq f_1(\ve)$. Taking the limit $\ve\to0$ in 
(\ref{pert}), we find 
\begin{equation}
\Phi(0)=2 N(\ve_F) V. \label{orderpert}
\end{equation}
As already mentioned, the exact answer is $\Phi(0)=\phi(\ve_F)/\pi$, with the
phase shift given by (\ref{phaseshift}). We see that the approximate answer
(\ref{orderpert}) is correct to first order in $V$, which is expected due
to our first order expansion of the potential in the bosonic fields.

\begin{figure}
\begin{center}
\includegraphics[width=.45\columnwidth]{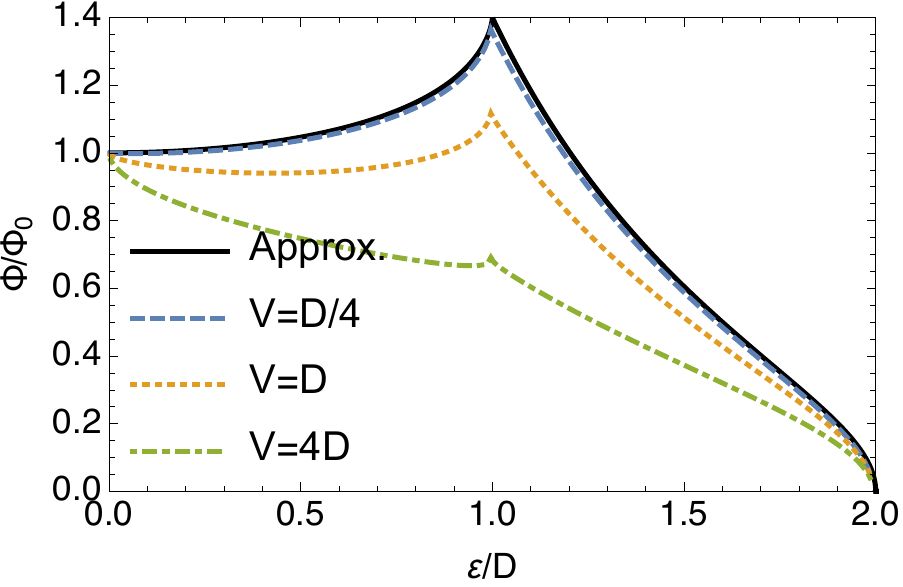}
\caption{Rescaled bosonic displacement $\Phi(\ve)/\Phi(0)$ defined in
Eq.~(\ref{opar}), for a cosine electronic band at half-filling. 
The solid curve represents the analytical result 
for weak interaction, $f_1(\ve)/f_1(0)$, with $f_1(\ve)$ given by (\ref{pert}), 
which is independent of $V$. The dashed, dotted and dot-dashed curves represent 
respectively exact results obtained by exact diagonalization of the original 
electronic model, for the various values $V=D/4,\,D,\,4D$. The analytical theory 
is accurate up to $V=D/4$ on all energy scales.
\label{f1}}
\end{center}
\end{figure}

The advantage in our method becomes more evident when investigating the full 
profile of the bosonic displacement $\Phi(\ve)$ as a function of $\ve$. In 
Figure \ref{f1} we compare approximate and exact 
results for $\Phi(\ve)/\Phi(0)$ in the case of a cosine dispersion relation
(namely for a nearest neighbor tight-binding band), for which
\begin{equation}
N(\ve)=\frac{\theta(D-|\ve|)}{2\pi D\sqrt{1-\left(\ve/D\right)^2}}.\label{cosband}
\end{equation}
We consider here half-filling, {\it i.e.} $\ve_F=0$. The exact results were obtained by
exact diagonalization of the fermionic Hamiltonian, for respectively $V=D/4,\,D$ and
$4D$. For $V<D/4$, the exact result is nearly indistinguishable from the
approximate result, for all energies $\ve$. We see that even when $V$ is so large that
the approximate result is no longer quantitatively accurate, there are still
strong qualitative similarities between the exact and approximate result.

Now we discuss the calculation of the transition rate $W(\ve)$ for the approximate 
ground state $\left|f_1\right>$ given in Eq.~(\ref{coherent}). 
According to Eq.~(\ref{pt}), we obtain $W(\ve)$ from the Fourier transform of the function
\begin{equation}
P(t)=\left<f_1\right|\tilde \psi_0^\dagger e^{-iH_0 t} \tilde \psi_0\left|f_1\right>.
\end{equation}
In the time representation, the above expression reads 
\begin{eqnarray}
P(t)&=&\left<f_1\right|\tilde \psi_0^\dagger e^{-iH_0 t} \tilde \psi_0\left|f_1\right>\nonumber\\
&=&4\pi\int_{-\infty}^\infty \!\!d\tau_1 \int_{-\infty}^\infty \!\!d\tau_2
\Delta(\tau_1)\Delta(\tau_2-t)^*
\left<f_1\right|\psi(\tau_1)^\dagger \psi(\tau_2) e^{-i H_0t}\left|f_1\right>.\label{pt3}
\end{eqnarray}
We have encountered the operator $\psi(\tau_1)^\dagger\psi(\tau_2)$ before in
the potential energy term of $H_V$, and we use its bosonized form. After some algebra,
we obtain
\begin{eqnarray}
\left<f_1\right|\psi(\tau_1)^\dagger &\psi(\tau_2) e^{-i H_0t}\left|f_1\right>=
\int_0^\infty \!\!\frac{d\omega}{2\pi}\,e^{i(\omega-\ve_F)(\tau_1-\tau_2)}\nonumber\\
&\times\left<f_1\right|e^{-i\left[\varphi^\dagger(\tau_1)-\varphi^\dagger(\tau_2)\right]}
e^{-i\left[\varphi(t_1)-\varphi(t_2)\right]}e^{-i t\int_0^\infty \!\!d\ve\,\ve b_\ve^\dagger b_\ve}\left|f_1\right>.
\end{eqnarray}
The expectation value is evaluated using standard coherent state technology. The
result is an analytical expression for $P(t)$ in terms of $f_1(\ve)$. It reads 
\begin{equation}
P(t)=2\exp\left[\int_0^\infty \!\!\frac{d\ve}{\ve}\,
f_1(\ve)^2\left(e^{-i\ve t}-1\right)\right]\int_0^\infty 
\!\!d\omega\, [G(\omega,t)]^2,\label{pt2}
\end{equation}
where
\begin{eqnarray}
\hspace{-2cm}
G(\omega,t)=\int_{-\infty}^\infty \!\!d\tau\, e^{-i(\omega-\ve_F)\tau} 
\Delta\left(\tau-\frac{t}{2}\right)^*
\exp\left[-2i\int_0^\infty \!\!
\frac{d\varepsilon}{\varepsilon}\,\sin(\varepsilon\tau)e^{-i\ve
t/2}f_1(\varepsilon)\right].
\label{gtw}
\end{eqnarray}
The above integrals can be implemented as fast Fourier transforms so that
the computation time for $P(t)$ scales like $\Omega \ln \Omega$ where $\Omega$
is the size of the energy grid used to discretize $f_1(\ve)$. 

Before addressing our result for a realistic band structure, we  show that our 
approach reproduces the standard bosonization results~\cite{Schotte,Giam} for 
a linearized spectrum. For this purpose, we replace the microscopic density of states
with an effective one in which modes far from the Fermi energy are suppressed by a soft cut-off:
\begin{equation}
N_{\rm eff}(\ve)=N(\ve_F)e^{-|\ve|/\Lambda}
\end{equation}
with $\Lambda\ll D$. 
The effective density of states should be thought of as the result of integrating out high energy modes,
and is therefore accompanied by a renormalization of the system parameters $V$ and $\gamma$. 
This leads to a displacement~(\ref{pert}),
$f_1(\ve)=2N(\ve_F)V e^{-\ve/2\Lambda}$. For $t\gg1/\Lambda$, this gives
\begin{equation}
P(t)=\frac{2^{1-2f_1(0)}N(\ve_F)}{it+0^+}\left(\frac{\Lambda}{0^++it}\right)^{[f_1(0)+1]^2-1},\label{ptflat}
\end{equation}
For $|\ve|\ll\Lambda$, the transition rate $W(\ve)$ is proportional to the 
Fourier transform of (\ref{ptflat}). By making a change of integration variable $t\to\ve t$ in the Fourier transform and 
noting the analiticity in the lower half of the complex $t$-plane of (\ref{ptflat}), one then readily obtains
\begin{equation}
W(\ve)=\theta(-\ve)W_\Lambda 
\left|\frac{\ve}{\Lambda}\right|^{\left[f_1(0)+1\right]^2-1},
\end{equation}
with some cut-off dependent constant $W_\Lambda$. 
As discussed previously, the microscopic power-law behavior at the threshold can 
be predicted by standard bosonization provided the exact phase shift is used 
in place of linearized expression $\pi f_1(0)=2\pi VN(\ve_F)$.
On the other hand, it is usually a daunting task to relate the prefactor $W_\Lambda$ to the bare 
system parameters within the standard bosonization method, as this requires explicitly 
working out how the system parameters are renormalized when ultraviolet modes 
are eliminated. The microscopic bosonization method can however get access to this 
prefactor, and in fact to the whole energy dependence of the transition rate,
as we demonstrate now.
\begin{figure*}[ht]
\begin{center}
\includegraphics[width=.45\textwidth]{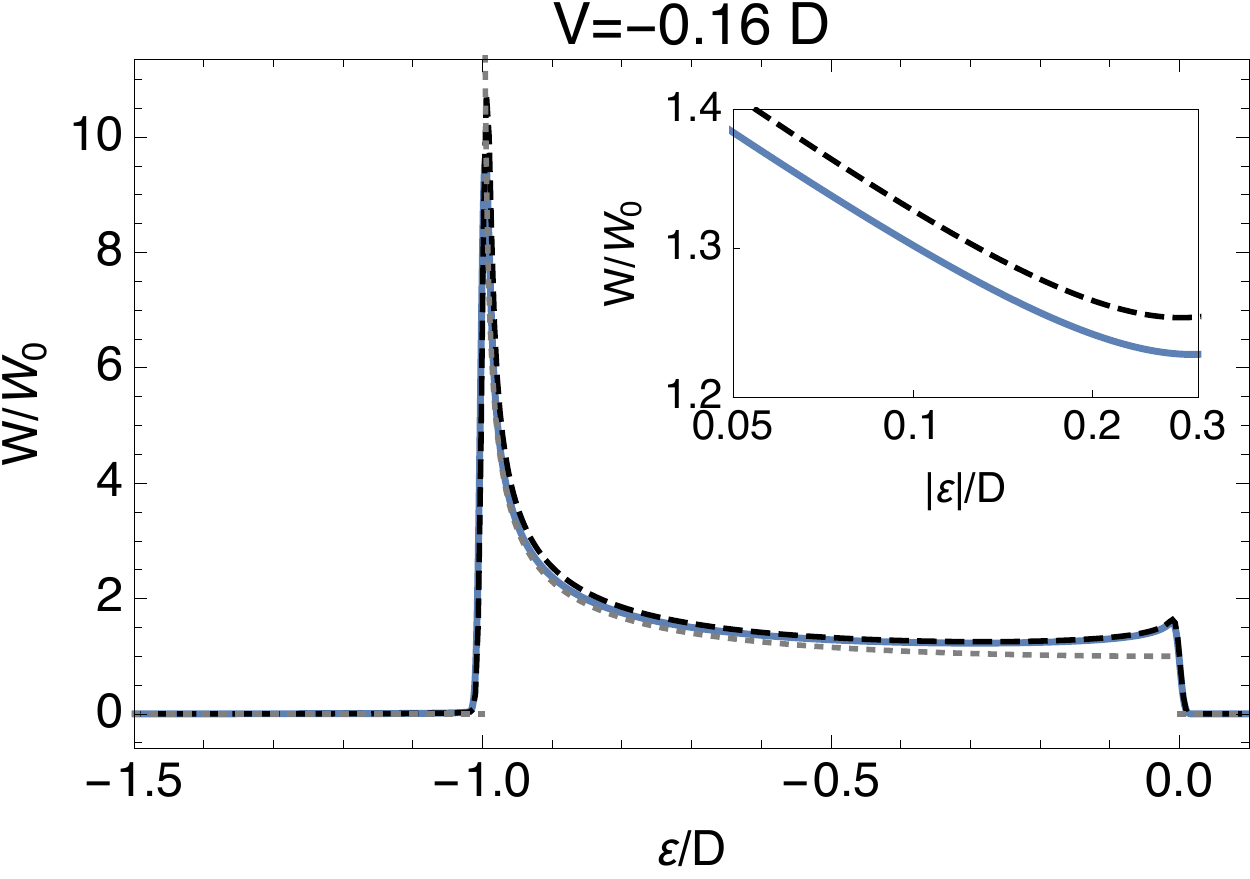}
\includegraphics[width=.45\textwidth]{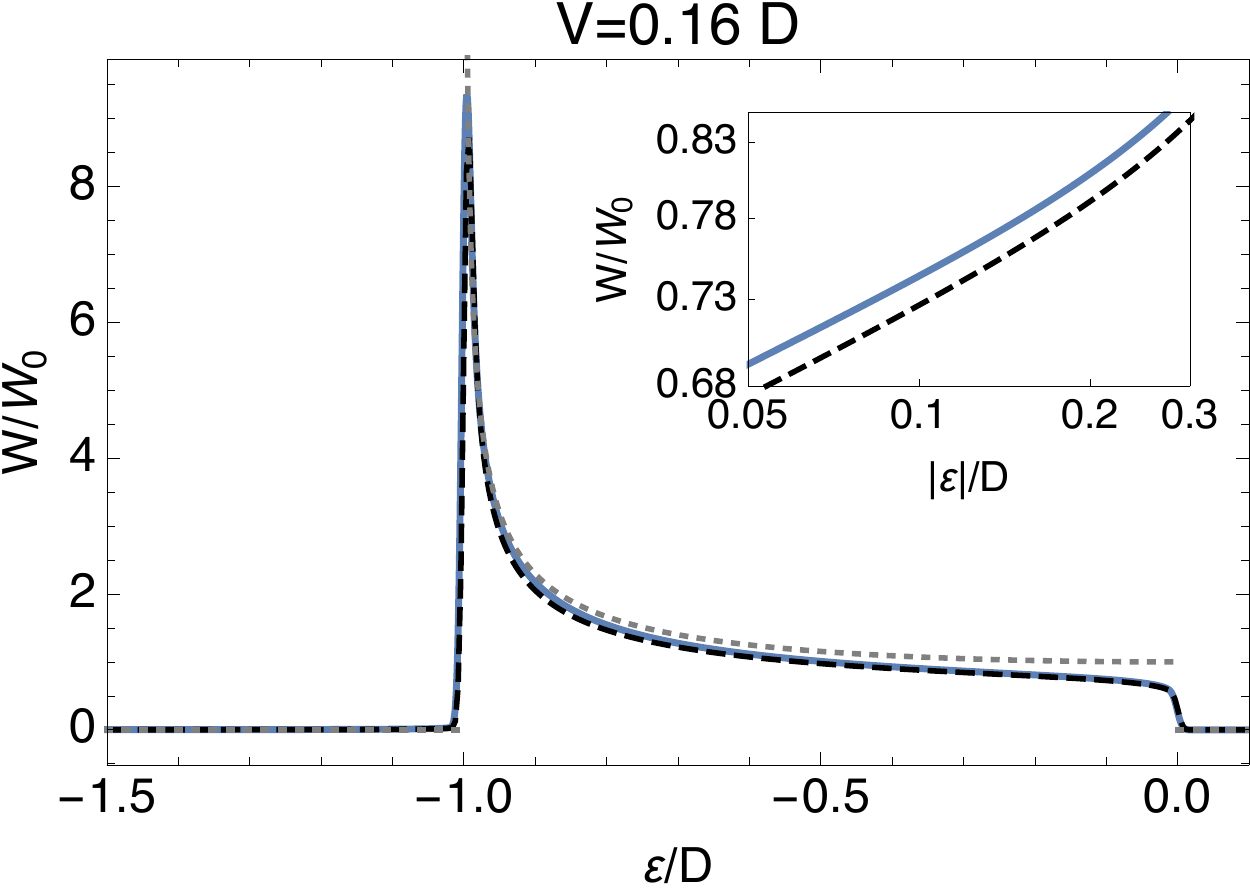}
\caption{X-ray transition rate $W(\ve)$ of Eq.~(\ref{pt}), in units of 
$W_0=\gamma^2/2\pi D$, for a half-filled cosine band. The two panels correspond
respectively to the small negative and positive values of the impurity interaction $V=\pm0.16D$, 
which leads to a phase shift $\phi=\pm 0.1 \times \pi/2$. Solid blue lines correspond to the
first order analytical formula (\ref{pt2}), dashed black lines show the exact 
numerical result obtained by exact diagonalization of the electronic problem, and dotted 
grey lines show the $V=0$ result, which is proportional to the free density of states. 
Insets show $W(\ve)$ vs. $|\ve|$ in log-log scale, in order to highlight the x-ray 
singularity at the threshold. 
\label{f2}}
\end{center}
\end{figure*}

For the cosine band~(\ref{cosband}), we have calculated the x-ray transiton rate $W(\ve)$ 
in the first order approximation (\ref{pt2}) of the bosonized theory, and compared to 
exact results obtained by direct diagonalization of the original fermion Hamiltonian.
(See Fig.~\ref{f2}.) 
We find excellent quantitative agreement for $|V|=\pm0.16D$ on all energy
scales, not only near the threshold, but also within and outside the band. The
van Hove singularity at the band bottom is also taken into account properly by
this simple analytical treatment. Because the first order expansion only captures the
phase shift correctly for $V\lesssim D/4$, the method cannot be trusted regarding the
x-ray emission spectrum for larger $V$ values.
A more accurate variational approach for, valid even for $V>D$ will now be 
developed in Sec.~\ref{sec5N}, and we defer to this section a thorough
discussion of the x-ray edge spectra.

\section{Variational bosonic solution of the microscopic x-ray edge problem}
\label{sec5N}
\subsection{Single coherent state variational theory}
\label{sec51}
For moderate interaction strengths $V<D/4$, the analytical approximation that we presented 
in the previous section was found to capture the physics of the x-ray problem quantitatively
on all energy scales. For larger interaction strengths, the approximation remained only 
qualitatively predictive, as seen in Fig.~\ref{f1}. This is already a remarkable success 
for a microscopic bosonization approach of electronic lattice models.
In the first part of this section we investigate a straightforward variational generalization
of the analytical theory that builds on the natural structure of the wave function in terms of
bosonic coherent states.
We will see that the variational Ansatz presented in this section 
captures important aspects of the x-ray edge physics at large $V$, 
especially beyond the band edge, but still leaves room for improvement. In the 
second part of this section, we will formulate an improved variational Ansatz
which produces a rate $W(\ve)$ that, though not exact, is very accurate up to phase shifts nearly 
equal to the maximal value $\pi/2$. This will serve to illustrate that a fully
microscopic bosonization calculation can be performed on tight-binding models even beyond
the weak interaction regime, an aspect that clearly clashes with the common wisdom about 
bosonization.

Our first variational Ansatz is simply a generic coherent state
\begin{equation}
\left|\Psi_{\rm var}\right>=\left|f\right> =
\exp\left\{-\int_0^\infty \!\!\frac{d\varepsilon}{\sqrt{\varepsilon}}\,
f(\varepsilon)\left[b_\varepsilon^\dagger- b_\varepsilon\right]\right\}
\left|{\rm vac}\right>
\label{restricted},
\end{equation}
which is obtained by promoting $f_1(\ve)$ in the approximate ground state (\ref{coherent}) 
to a function $f(\ve)$ that is determined by minimizing the energy 
$E_{\rm var}=\left<f\right|H_{V,{\rm enl}}\left|f\right>$. Although the
computations are performed entirely within the bosonized theory, we provide in
\ref{meaning} a physical interpretation of this coherent state in terms of the
original fermions.
This coherent state Ansatz is guaranteed to produce a Fermi edge singularity of
the form (\ref{powerlaw1}), with a phase shift $\phi(\varepsilon_F)=\pi f(0)$.
Our main goal in using this variational state is to account for the
non-linear behavior of the phase shift as a function of $V$, Eq.~(\ref{phaseshift}), 
which lies beyond the leading-order expression~(\ref{orderpert}).
One can verify that $\pi f(0)$ corresponds to the phase shift at
the Fermi level, by studying the overlap with the vacuum state $|{\rm
vac}\rangle$:
\begin{equation}
\olap{\rm vac}{f}=\exp\left[-\frac{1}{2}\int_0^\infty \!\!d\ve\,\frac{f(\ve)^2}{\ve}\right],
\end{equation}
which vanishes because the integral in the exponent is logarithmically divergent
at small $\ve$. In a finite system, the divergence is cut off at an energy $\sim
v_F/L$, where $L$ is the system size. The overlap $\olap{\rm vac}{f}$ thus
vanishes like $L^{-f(0)^2/2}$. In view of Anderson's orthogonality
theorem~\cite{PWAcatastrophe}, we again identify $\pi f(0)$ as the phase shift 
at the Fermi energy.

For the single coherent state Ansatz, the energy functional takes the
following explicit form:
\begin{equation}
E_{\rm var}=\int_0^\infty \!\!d\varepsilon\, f(\varepsilon)^2
+2V\int_0^\infty \!\!d\omega\, A(\omega)^2-\frac{V}{2}, 
\end{equation}
where
\begin{equation}
A(\omega)=\int_{-\infty}^\infty \!\!d\tau\, e^{-i(\omega-\ve_F)\tau} 
\Delta(\tau)^*e^{-2i\int_0^\infty \!\!\frac{d\varepsilon}{\varepsilon}\,
\sin(\varepsilon\tau)f(\varepsilon)},\label{aomega}
\end{equation}
is real.
The functional derivative with respect to $f(\ve)$ is
\begin{eqnarray}
\frac{\delta E_{\rm var}}{\delta f(\varepsilon)}
&=&2f(\varepsilon)+\frac{4V}{\varepsilon}
\int_0^\infty \!\!d\omega\, A(\omega)\left[A(\omega+\varepsilon)-A(\omega-\varepsilon)\right]\nonumber\\
&=&2f(\ve)-\frac{4iV}{\pi\ve}\int_{-\infty}^\infty \!\!d\tau\,\sin(\ve\tau)A_1(-\tau)A_2(\tau),\label{dHdf}
\end{eqnarray}
where
\begin{eqnarray}
A_1(\tau)&=&\int_0^\infty \!\!d\omega\,A(\omega)e^{i\omega\tau},\\
A_2(\tau)&=&\int_{-\infty}^\infty \!\!d\omega\,A(\omega)e^{i\omega\tau}=2\pi \Delta(\tau)^*e^{i\ve_F\tau-2i\int_0^\infty \!\!
\frac{d\varepsilon}{\varepsilon}\,\sin(\varepsilon\tau)f(\varepsilon)}.\label{a1a2}
\end{eqnarray}
The final lines of (\ref{dHdf}) and (\ref{a1a2}) are convenient for numerical
calculations, because all integral transforms that are involved can be
implemented as fast Fourier transforms. The computation time for calculating
$E_{\rm var}$ and its gradient again scales like $\Omega \ln \Omega$, where
$\Omega$ is the size of the energy grid used to discretize $f(\ve)$. Note that
(\ref{dHdf}) can be solved for $f(\ve)$ to linear order in $V$, by setting
$f(\ve)=0$ in $A(\omega)$. By doing so, we recover $f(\ve)=f_1(\ve)$, where
$f_1(\ve)$ is given by the first order analytical formula~(\ref{pert}). Thus, 
at small $V$, the single coherent state variational Ansatz reduces to our 
previous approximation where the potential is expanded to first order in 
boson operators.
Once the variational state is optimized, we need to calculate the correlation function
$P(t)$ in Eq.~(\ref{pt}), and from there, the transition rate $W(\ve)$. Since the variational
state has the same coherent state form as the analytical approximation of the previous section,
we can do so simply by replacing $f_1(\ve)$ by $f(\ve)$ in the equations
(\ref{pt2}) and (\ref{gtw}) of the previous section.
The details of our numerical implementation of the variational calculation can
be found in the supplementary material. We were comfortably able to perform the
variational calculation at an energy resolution of $10^{-3}D$.

The results we present are for a cosine band (\ref{cosband}), 
leading to the explicit form of
\begin{equation}
\Delta(\tau)=\frac{\Gamma(3/4)J_{1/4}(|D\tau|)}{2\pi |2D\tau|^{1/4}},
\end{equation}
where $\Gamma(z)$ is the Gamma function, and $J_\nu(z)$ is the Bessel function of order $\nu$.
Because of particle-hole symmetry, {\it i.e.} $N(\ve)=N(-\ve)$, $\Delta(\tau)$ 
is real and even. The Fermi energy is set in the middle of the band throughout, 
and the phase shift is given exactly by:
\begin{equation}
\phi(\ve_F=0)=\arctan(V/D).\label{vphi}
\end{equation}
\begin{figure*}
\begin{center}
\includegraphics[width=.45\textwidth]{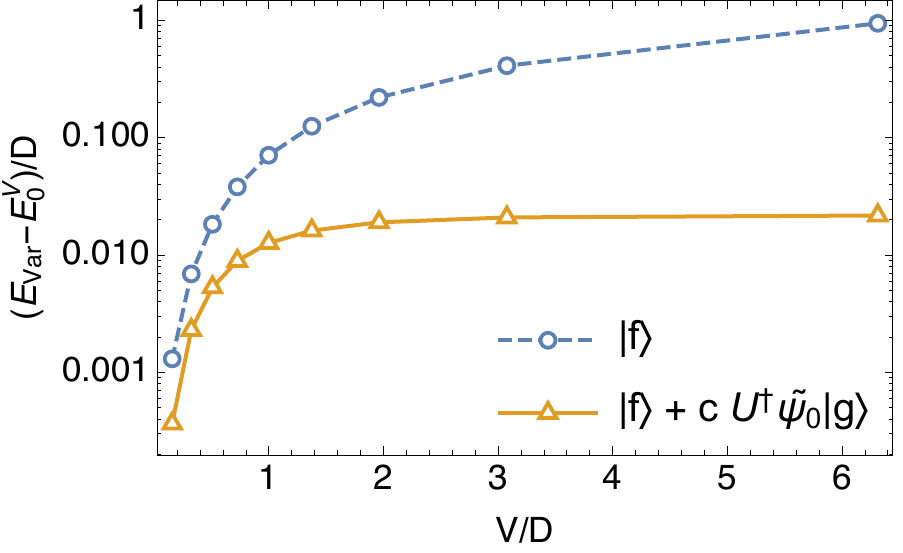} ~
\includegraphics[width=.45\textwidth]{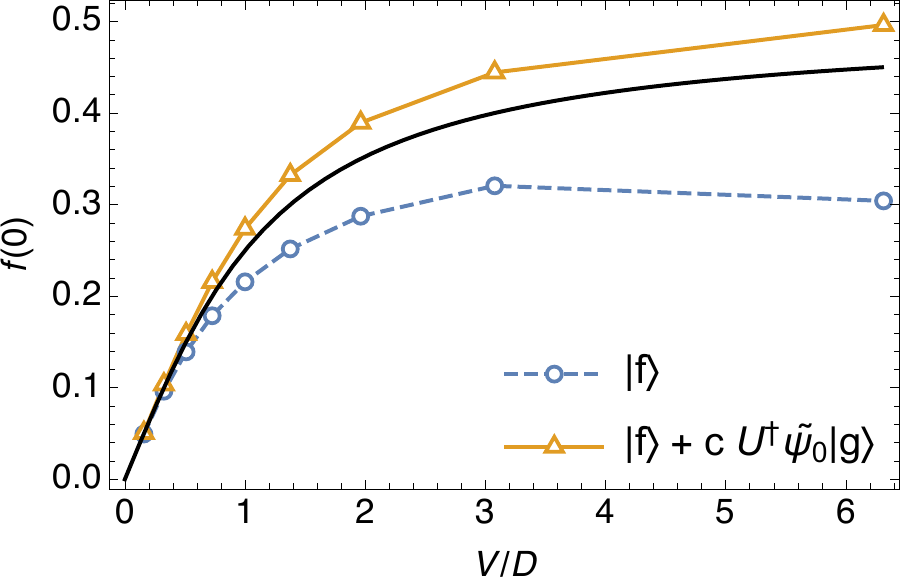} 
\end{center}
\caption{The left panel shows the variational energy $E_{\rm var}$ relative
to the exact ground state energy $E_0^V$ of $H_V$, versus potential strength,
both for the single coherent state Ansatz (dashed blue line with circles)  $|f\rangle$ of
Eq.~(\ref{restricted}) and the improved superposed Ansatz (solid orange line with 
triangles) $|f\rangle+cU^\dagger\tilde\psi_0|g\rangle$ of Eq.~(\ref{full_ansatz}), to 
be discussed in Sec. \ref{sec52}.
In the right panel, the variational parameter $f(0)$ corresponding to the phase
shift (divided by $\pi$) at the Fermi energy for both the single coherent state and
the superposed Ansatz is similarly compared to the exact phase shift (solid black line). 
\label{f3}} 
\end{figure*}
In the left panel of Figure \ref{f3}, we compare the minimized variational
energy $E_{\rm var}$ for various $V$ to the true ground state energy of the
infinite system, which is given by 
\begin{equation}
E_0^V=D\left(1-\sqrt{1+(V/D)^2}\right)/2. 
\end{equation} 
At the smallest potential strength that we considered, namely $V=0.16D$, 
the coherent state Ansatz yields an energy that is accurate to within the 
discretization error $\sim 10^{-3} D$.
However, the error $E_{\rm var}-E_0^V$ grows to a significant fraction of the
band width $2D$ when $V$ becomes large, indicating that the Ansatz does not
provide a quantitatively accurate description of the ultraviolet modes that are
affected by the potential (this problem will be cured by an improved Ansatz in
what follows).

In the right panel of Figure \ref{f3}, we plot the variational parameter $f(0)$
as a function of $V$, which gives $\pi^{-1}$ times the phase shift of fermions
at the Fermi energy, for the variational state. We compare it to the exact
phase shift (also divided by $\pi$), given by $\arctan(V/D)/\pi$, when the Fermi
energy is in the middle of the band. From (\ref{pert}) it follows that to
lowest order in $V$, the variational state reproduces the correct phase shift
$2\pi N(\ve_F) V$, as we can check on the plot.  At large positive $V$, 
the single coherent state Ansatz nicely corrects for the unbounded growth found 
at large $V$ in Schotte and Schotte's solution, although it significantly underestimates 
the exact phase shift.
The variational Ansatz respects the Friedel sum rule, so that the average number 
of particles displaced by the potential energy term is $f(0)$. The underestimation 
of $f(0)$ at large positive $V$ therefore implies that the single coherent state 
Ansatz displaces too few particles at large $V$.

We now use the optimized coherent state trial wave function to compute the x-ray 
transition rate $W(\ve)$ and compare the variationally obtained rate to numerically 
exact diagonalization results. (See Supplementary Material for details about the implementation.)
Before discussing the variational results, it is useful to highlight the
following features of the numerically exact results. As $\ve$ approaches the
threshold $\ve=0$ from below, $W(\ve)$ displays power-law behavior (the Fermi edge
singularity) with an exponent in agreement with the analytical prediction
$[\phi(0)/\pi+1]^2-1$, where $\phi(0)$ is given by (\ref{vphi}). In the $V=0$
limit, we know that $W(\ve)$ diverges as $1/\sqrt{1+\ve/D}$ when $\ve$
approaches $-D$ from above, due to the van Hove singularity in $N(\ve)$. When
$V$ is varied, this peak does not seem to significantly broaden in our exact
diagonalization data, and we conclude that the van Hove divergence remains present 
when $V\not=0$. 
For $V=0$, the transition rate is strictly zero below the band bottom for $\ve<-D$. 
However, the rate develops a tail in the region $\ve<-D$ for non-zero $V$.
Since $\ve<-D$ corresponds to more excitation energy than a single particle-hole pair can carry,
the tail is necessarily associated with multiple-particle excitations. It seems 
from our numerically exact data that $W(\ve)$ tends to a finite value as $\ve$ 
approaches $-D$ from below. 

\begin{figure*}
\begin{center}
\includegraphics[width=.45\textwidth]{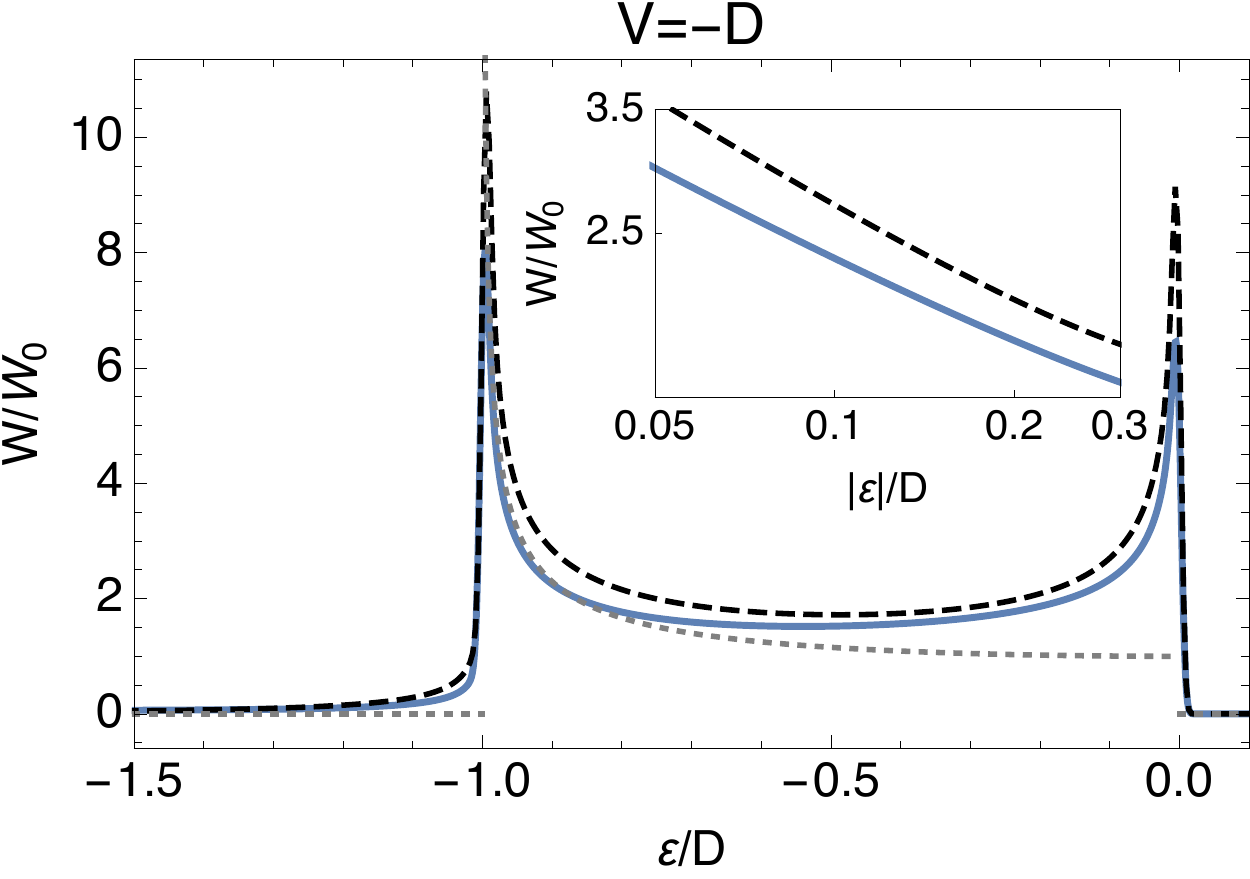}
\includegraphics[width=.45\textwidth]{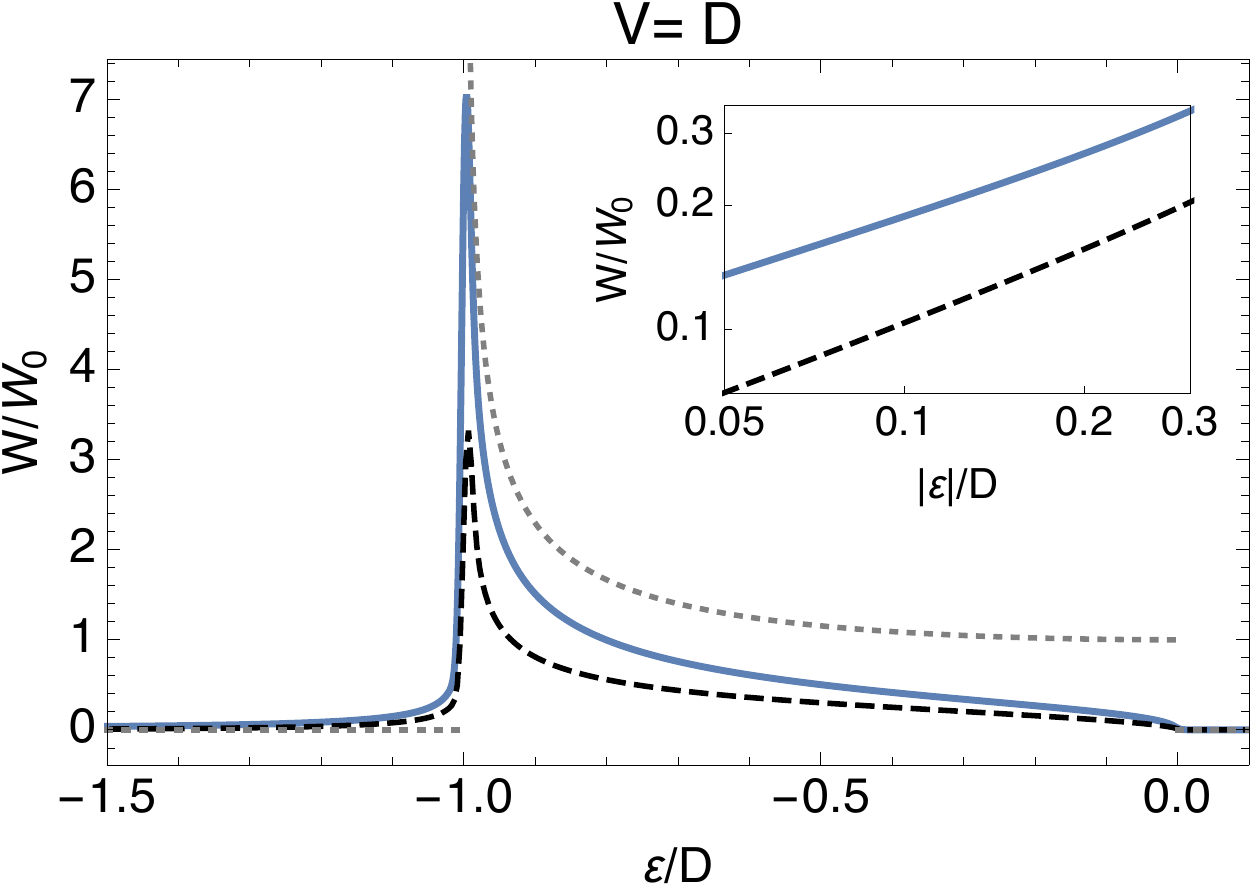}
\caption{X-ray transition rate $W(\ve)$ of Eq.~(\ref{pt}), in units of
$W_0=\gamma^2/2\pi D$, for a half-filled cosine band. The two panels correspond
respectively to the intermediate negative and positive values of the impurity
interaction $V=\pm D$, which leads to a phase shift $\phi=\pm 0.5 \times \pi/2$.
Solid blue lines correspond to the single coherent state Ansatz (\ref{restricted}), 
dashed black lines show the exact numerical result, and dotted grey lines show the 
$V=0$ result, which is proportional to the free density of states. Insets 
show $W(\ve)$ vs. $|\ve|$ in log-log scale, in order to highlight the x-ray singularity 
at the threshold.
\label{f4}}
\end{center}
\end{figure*}

For $|V|\lesssim D/4$, the variational results are nearly identical to  
to the analytical calculation of Sec.~\ref{sec4N}, and thus are quantitatively 
accurate on all energy scales.
Figure~\ref{f4} demonstrates more-than-qualitative agreement of the transition
rate at moderately large values of the potential $V=\pm D$ for the
single coherent state Ansatz~(\ref{restricted}). At both positive and negative $V$, we find
that the variational results capture both the infrared physics of the Fermi
edge singularity close to $\ve=0$, and the band structure physics of the van
Hove singularity at $\ve=-D$. Remarkably, the tail beyond the band edge at $\ve<-D$ 
is also reproduced.  At large negative $V$, deviations from the exact result
become more pronounced, but the general shape and scale of the variational curve is still 
similar to the exact result. For instance, the deviation
$\left|1-W_{\rm var}(\ve)/W_{\rm exact}(\ve)\right|$,
averaged over the interval $\ve\in(-D,0)$, peaks at $\sim 15\%$ as $|V|$ is
increased, and the largest contribution to the error comes from the vicinity of
the singularities at $\ve=0$ and $\ve=-D$. 
At large positive $V$, we find a more important mismatch to the exact results, 
with the variationally calculated rate significantly larger than the true
one. Curiously, the shape of the variationally determined transition rate remains in
excellent agreement with the exact answer. For instance, at $V=D$, scaling the 
variational rate by $0.55$ produces a result nearly identical to the exact result 
for all $\ve$ (not shown).

According to the definition of the transition rate~(\ref{rate}), one finds the
sum rule:
\begin{equation}
\int_{-\infty}^0 \!\!d\ve\, 
W(\ve)=\gamma^2 \left<\tilde \psi_0^\dagger \tilde \psi_0\right>,
\end{equation}
{\it i.e.} the area under the curve of $W(\ve)$ is proportional to the average number
of particles at the impurity site. The fact that the single coherent state Ansatz
significantly overestimates the transition rate $W(\ve)$ at large positive $V$, implies that it predicts
the wrong average number of particles on the site zero of the lattice.
Yet, it is quite surprising that the overall line-shape of the emission
spectrum is so well described. This useful piece of information will lead to a
drastic improvement of the variational Ansatz, that we consider next.

\subsection{Improved variational Ansatz as superposed coherent states}
\label{sec52}

We now propose an improved Ansatz, based on the previous considerations.
We found that the single coherent state Ansatz~(\ref{restricted}) produces an
x-ray transition rate $W(\ve)$ that is qualitatively correct up to large negative 
$V$, and that has almost the perfect line shape at large positive $V$ (but not the
correct scale overall). Owing to the form of the correlation function~(\ref{pt})
involved in the transition rate, the component of the wave function in which unit cell zero
is empty will not affect the shape of the transition rate, only its magnitude.
For $V>0$, we therefore consider an Ansatz of the form
\begin{equation}
\left|\Psi_{\rm var}\right>=\left|f\right>+c\, U^\dagger \tilde \psi_0 \left|g \right>.
\label{full_ansatz}
\end{equation}
Here $\left|f\right>$ and $\left|g\right>$ are single coherent states 
(\ref{coherent}), with real functions $f(\ve)$, $g(\ve)$ and a real
relative weight $c$, all to be optimized variationally.
Since the trial state consists of two terms with distinct configurations of collective
degrees of freedom, we refer to it as the superposed Ansatz. It is interesting
to note that the second term in the Ansatz can be written in the form
$\int_{-\infty}^\infty d\tau\, K(\tau) \left|g_\tau\right>$ where
$\left|g_\tau\right>$ is a coherent state with displacement
$g_\tau(\ve)=g(\ve)+\exp[-(i\tau+a/2)\ve]$. Of course, an arbitrary state can be
written as a superposition of coherent states. However, in the general case, 
the exact decomposition consists of
an infinite-dimensional integral, with two dimensions for every bosonic
mode, because each mode displacement can vary independently
over the whole complex plane. Here in contrast, 
we are dealing with a single one-dimensional integral,
that limits the the degree of entanglement in the trial state.
A similar sparseness was encountered previously in the systematic coherent state 
expansion that the authors developed to deal with Kondo-type impurity models in 
an infinite flat band~\cite{Bera,Snyman}. There, in fact, a handful of terms sufficed
to account very accurately for the proper many-body ground state.

For $c=0$, the superposed Ansatz reduces to the single coherent state trial
wave function of the previous section, but additional control over the occupation 
of the impurity site is provided by the second term, where the operator 
$U^\dagger\tilde \psi_0$ depletes unit cell zero of particles, without removing 
any particles from the band as a whole. 
For negative $V$, $U^\dagger \tilde \psi_0$ in the second term of
(\ref{full_ansatz}) is replaced by $U \tilde \psi_0^\dagger$. It is also
convenient to reverse the signs of $f(\ve)$ and $g(\ve)$, so that the negative
$V$ Ansatz reads \begin{equation} \left|\Psi_{\rm
var}\right>=\left|-f\right>+c\, U \tilde \psi_0^\dagger \left|-g
\right>.\label{eq2} \end{equation} Under particle-hole interchange, $b_\ve$ maps
onto $-b_\ve$. If the band possesses particle-hole symmetry, {\it i.e.} if
$N(\ve)=N(-\ve)$, the same optimal values of $c$, $f(\varepsilon)$ and
$g(\varepsilon)$ then minimize the energy at $(\ve_F,V)$ and $(-\ve_F,-V)$.

Using the same arguments as before, one readily deduces that the second term in
the superposed Ansatz (\ref{full_ansatz}) is associated with a phase shift
$\phi(\ve_F)=\pi[g(0)+1]$. When $g(0)\not = f(0)-1$, all cross-terms in the
expectation value of the Hamiltonian vanish, owing to the Anderson orthogonality
theorem. As a result, one must have
\begin{equation}
g(0)=f(0)-1,
\end{equation}
in the optimized Ansatz for the energy actually to be lowered.
Note that unlike the single coherent state Ansatz, the superposed
Ansatz~(\ref{full_ansatz}) has to be normalized by hand.
Expressions for the associated energy functional and its gradient, in terms of
the variational parameters, can be found in the supplementary material. Although
lengthier than the expressions for the single coherent state Ansatz of
Sec.~\ref{sec5N}, they hold the same advantage over the exact diagonalization approach, 
namely that they can be implemented using a sequence of fast Fourier transforms, 
and the scaling of execution time with energy grid size is again linear 
(up to logarithmic corrections), in contrast to the cubic cost for the
exact diagonalization method.

For the superposed Ansatz~(\ref{full_ansatz}), it is easier to calculate 
$P(t)$ and hence the transition rate $W(\ve)$ for positive $V$ than for negative 
$V$. For positive $V$, the second term on the right hand side of (\ref{full_ansatz}) 
only contributes to $P(t)$ through a normalization constant, because $U^\dagger$ 
commutes with $\tilde \psi_0$ and $\tilde \psi_0^2=0$, so that \begin{equation}
P(t)=\frac{\left<f\right|\tilde \psi_0^\dagger e^{-iH_0 t}\tilde \psi_0\left|
f\right>}{\left<\Psi_{\rm var}\right|\left.\Psi_{\rm var}\right>}.
\end{equation} The numerator equals the right hand side of the expression
(\ref{pt2}) that we derived for $P(t)$ in the context of the analytical approximation of Sec.~\ref{ansol}. 
The denominator, which we also have to calculate when we
minimize $E_{\rm var}$, can be constructed from the expressions presented in
the supplementary material.
In order to calculate $P(t)$ at negative $V$ on the other hand, we must evaluate
the expectation value \begin{equation} \left<\Psi_{\rm var}\right|\tilde
\psi_0^\dagger e^{-iH_0 t}\tilde \psi_0\left|\Psi_{\rm var}\right>,
\end{equation} which contains non-vanishing cross-terms between the first and
second terms of the Ansatz, because compared to the $V>0$ case, products like
$\tilde \psi_0^2=0$ are replaced by $\tilde \psi_0^\dagger \tilde \psi_0\not=0$.
We have not been able to write the resulting expressions in such a way that
integral transforms can be implemented as fast Fourier transforms. 
For this reason, the results for $W(\ve)$ that we present in the next section are
restricted to positive $V$.

We discuss now our results for a cosine dispersion relation with the Fermi energy in the 
middle of the band. In the previous Figure \ref{f3} we compare both the minimized energy $E_{\rm var}$ 
and the phase shift $f(0)=\phi(\ve_F=0)/\pi$ of the superposed Ansatz to the exact
results for an infinite system and to the single coherent state Ansatz.
For the superposed Ansatz, in contrast to the single coherent state Ansatz, 
the error $E_{\rm var}-E_0^V$ saturates to $\sim 1\%$ of $D$ at large $V$, suggesting that 
ultraviolet modes are now accurately accounted for. For the rescaled phase shift $f(0)$, 
the superposed Ansatz~(\ref{full_ansatz}) also yields a significant improvement over the
single coherent state solution. In view of this success of the superposed 
Ansatz~(\ref{full_ansatz}), it is worth studying the 
lowest energy configuration of the bosonic degrees of freedom further. To this end, we
consider again the rescaled bosonic displacement (\ref{opar}), defined in
Sec.~\ref{sec2N}. 
For the single coherent state Ansatz, we obviously have $\Phi(\ve)=f(\ve)$. 
For the superposed Ansatz, the relationship between $\Phi(\ve)$ and the variational parameters 
is more complicated, and reads for a particle-hole symmetric band at half-filling:
\begin{equation}
\fl\Phi(\ve)=\frac{f(\ve)+c\left\{\left[f(\ve)+g(\ve)\right]B(0)+B(\ve)\right\}
+2c^2\int_0^\infty \!\!d\omega\,C(\omega)\left[C(\omega+\ve)+g(\ve)\right]}
{1+2cB(0)+2c^2\int_0^\infty \!\!d\omega\,C(\omega)^2},
\end{equation}
where
\begin{eqnarray}
\fl B(\omega)=\sqrt{2\Lambda}\int_{-\infty}^\infty \!\!d\tau\, e^{-i\omega\tau} \Delta(\tau)
e^{-\frac{1}{2}\int_0^\infty \!\!\frac{d\varepsilon}{\varepsilon}
\left\{[g(\varepsilon)-f(\varepsilon)]^2+2[g(\varepsilon)e^{i\varepsilon\tau}-f(\varepsilon)
e^{-i\varepsilon\tau}]+e^{-\varepsilon/\Lambda} \right\} },\\
\fl C(\omega)=\int_{-\infty}^\infty \!\!d\tau\, e^{-i\omega\tau} 
\Delta(\tau)e^{-2i\int_0^\infty \!\!\frac{d\varepsilon}{\varepsilon}\,
\sin(\varepsilon\tau)g(\varepsilon)}.
\end{eqnarray}
The function $B$ seems to depend on an as yet undefined parameter $\Lambda$ with dimensions 
of energy. However, this is not the case. The $\Lambda$-dependence of the
pre-factor is exactly cancelled by the $\Lambda$-dependence of the argument of
the exponential. Note that for the superposed Ansatz, it still holds that 
$\Phi(0)=f(0)$, {\it i.e.} $\Phi(0)$ corresponds to the average number of fermions 
displaced by the impurity potential.

\begin{figure*}[ht]
\begin{center}
\includegraphics[width=.45\textwidth]{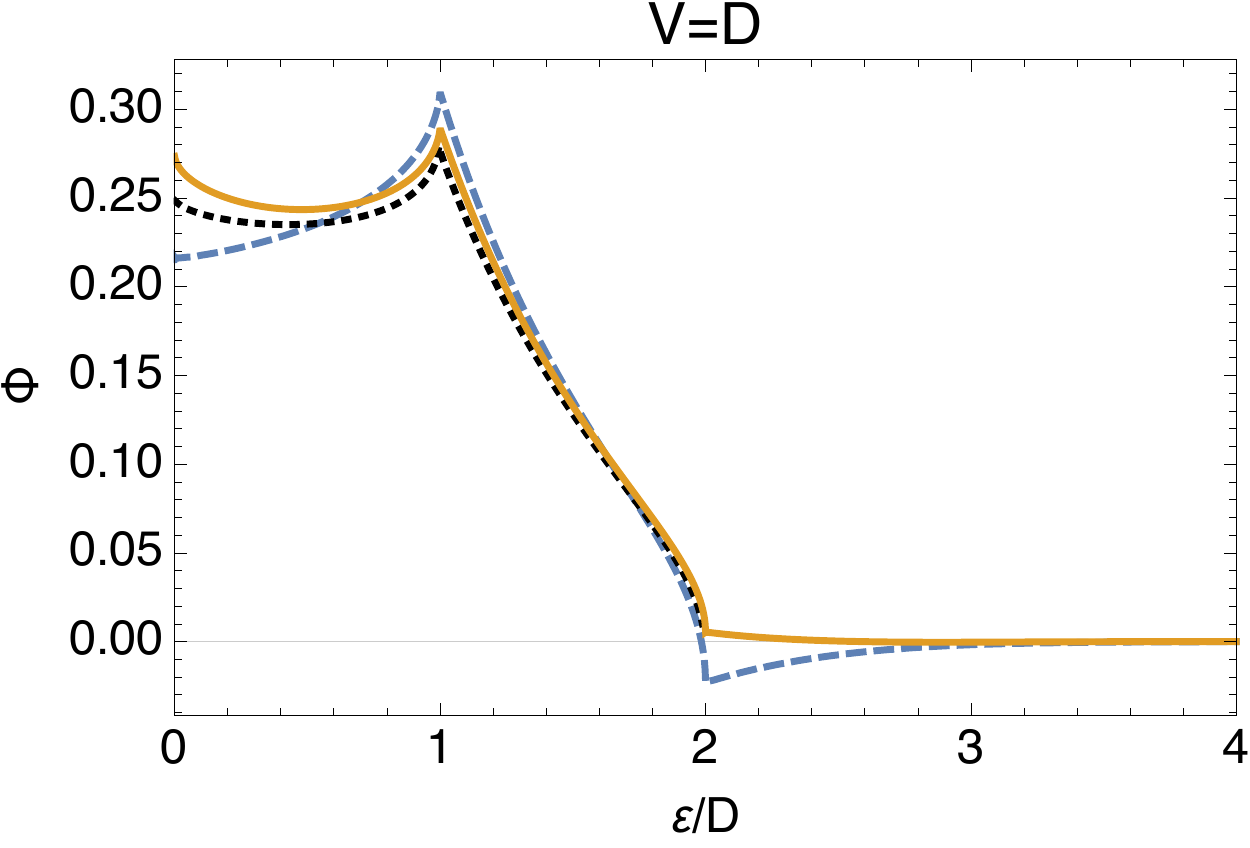}
\includegraphics[width=.45\textwidth]{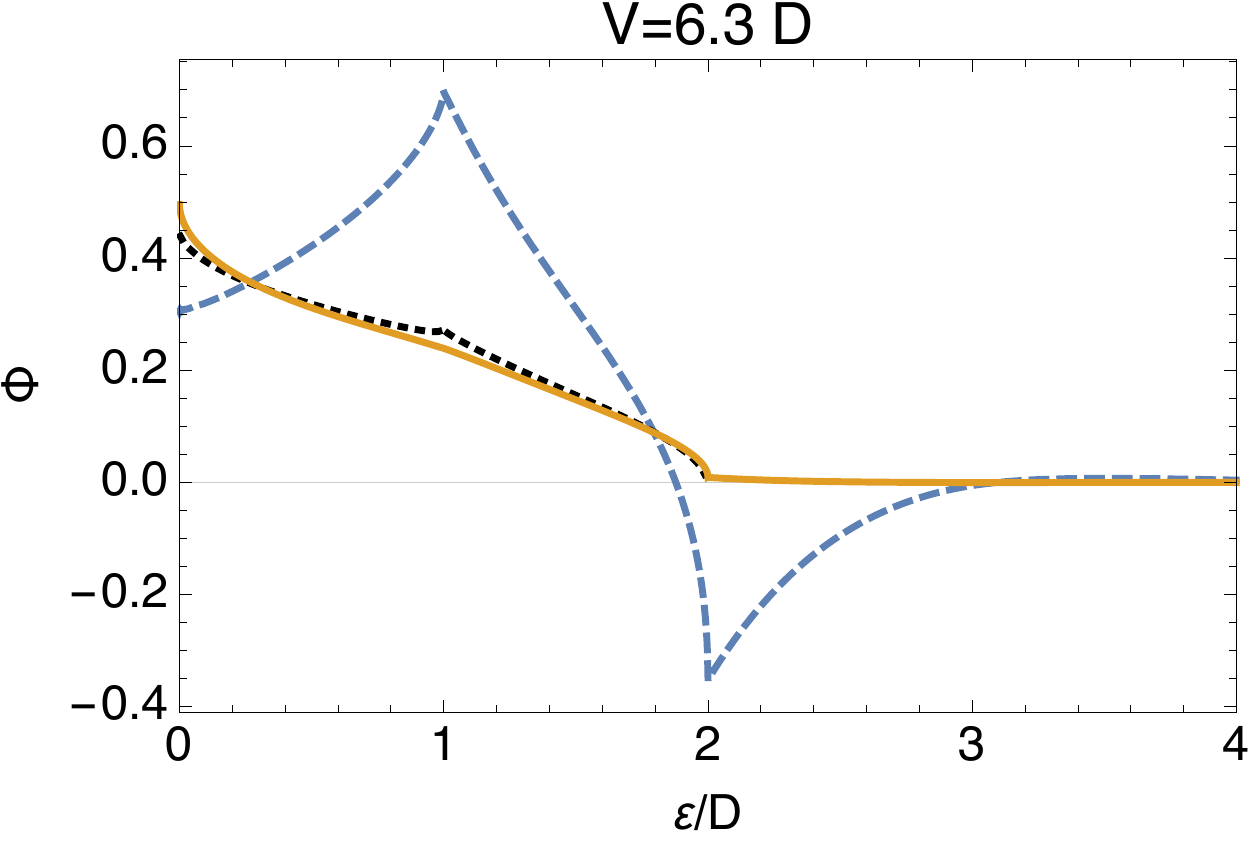}
\caption{The bosonic displacement $\Phi(\ve)$ of Eq.~(\ref{opar}) for two large values 
of the potential $V=D$ and $V=6.3D$, comparing the exact result (dotted black lines), 
the single coherent Ansatz $|f\rangle$ (dashed blue lines), and the superposed Ansatz
$|f\rangle+cU^\dagger\tilde\psi_0|g\rangle$ (solid orange lines). All numerical results 
are for a cosine band of width $2D$, at half filling.
\label{f5}}
\end{center}
\end{figure*}
In Figure \ref{f5}, we plot the exact displacement $\Phi(\ve)$ together with its 
estimated value according to the two variational Ans\"atze, for two large positive $V$
values (for negative $V$, the vertical axis is simply inverted). The displacement
$\Phi(\ve)$ shows several interesting and general features. At $\ve=0$ it equals 
$\phi/\pi$, the average number of particles displaced by the potential. In addition, 
$\Phi(\ve)$ has cusps at $\ve=D$ and at $\ve=2D$ which are related to the Friedel 
oscillations of the fermion density --- despite dispersive effects, the Fourier transform 
of $\Phi(\ve)$ roughly corresponds to the average particle density profile. The cusp value 
$\Phi(D)$ reaches a maximum at $V\sim 1.4D$ before decreasing again. This mirrors the amplitude 
of Friedel oscillations: Obviously, the amplitude is zero at $V=0$. The amplitude is also 
zero at $V\to \infty$, where the impurity cuts the crystal into two uncoupled semi-infinite
sections. In between these two limits, the amplitude first increases and then decreases. 
Finally, the displacement $\Phi(\ve)$ strictly vanishes for $\ve>2D$, because it is
associated to particle-hole excitations within a finite band.
At larger $V$, the single coherent state Ansatz
underestimates $\Phi(0)$, and overestimates $\Phi(D)$, while it is not well
suppressed at $\ve>2D$. Indeed the single coherent state does not allow enough 
freedom to entirely prevent the transfer of particles between band and spectator orbitals. 
In contrast, the spurious tail for $\ve>2D$ is very small in the superposed Ansatz, because 
the spectator excitations above the vacuum are energetically expensive and the superposed Ansatz 
allows for enough freedom to adjust ultraviolet modes optimally.

Globally, the superposed Ansatz agrees very well with the exact result for all $V$ and at all energies. 
The largest disagreement is found at small $\ve$ where, as we have seen in Figure \ref{f3}, 
the superposed Ansatz slightly overestimates the total displaced charge. At very large $V$, the 
superposed Ansatz does underestimate the strength of the kink in $\Phi(\ve)$
at $\ve=D$. Nonetheless, the non-monotone behaviour of $\Phi(D)$ as a 
function of $V$ is well reproduced.
Apart from elucidating how the superposed Ansatz~(\ref{full_ansatz}) improves
on the single coherent trial state, the above analysis also demonstrates that the 
bosonic description is less of a black box than the exact diagonalization method. It 
is possible to form an intuitive understanding of physical properties of the ground state, 
not only in the infrared, but at all scales, by inspecting the configuration of bosonic
degrees of freedom.

\begin{figure*}
\begin{center}
\includegraphics[width=.45\textwidth]{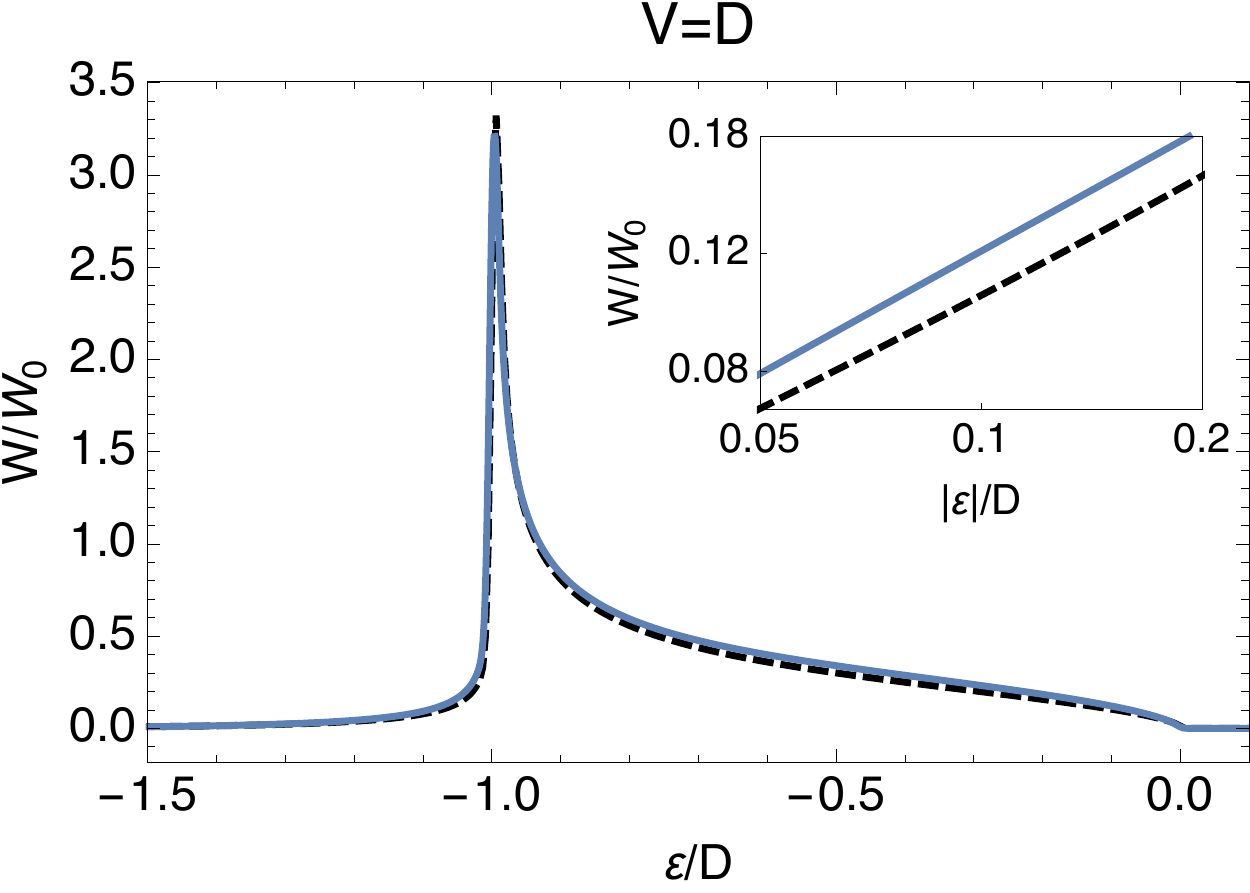}
\includegraphics[width=.45\textwidth]{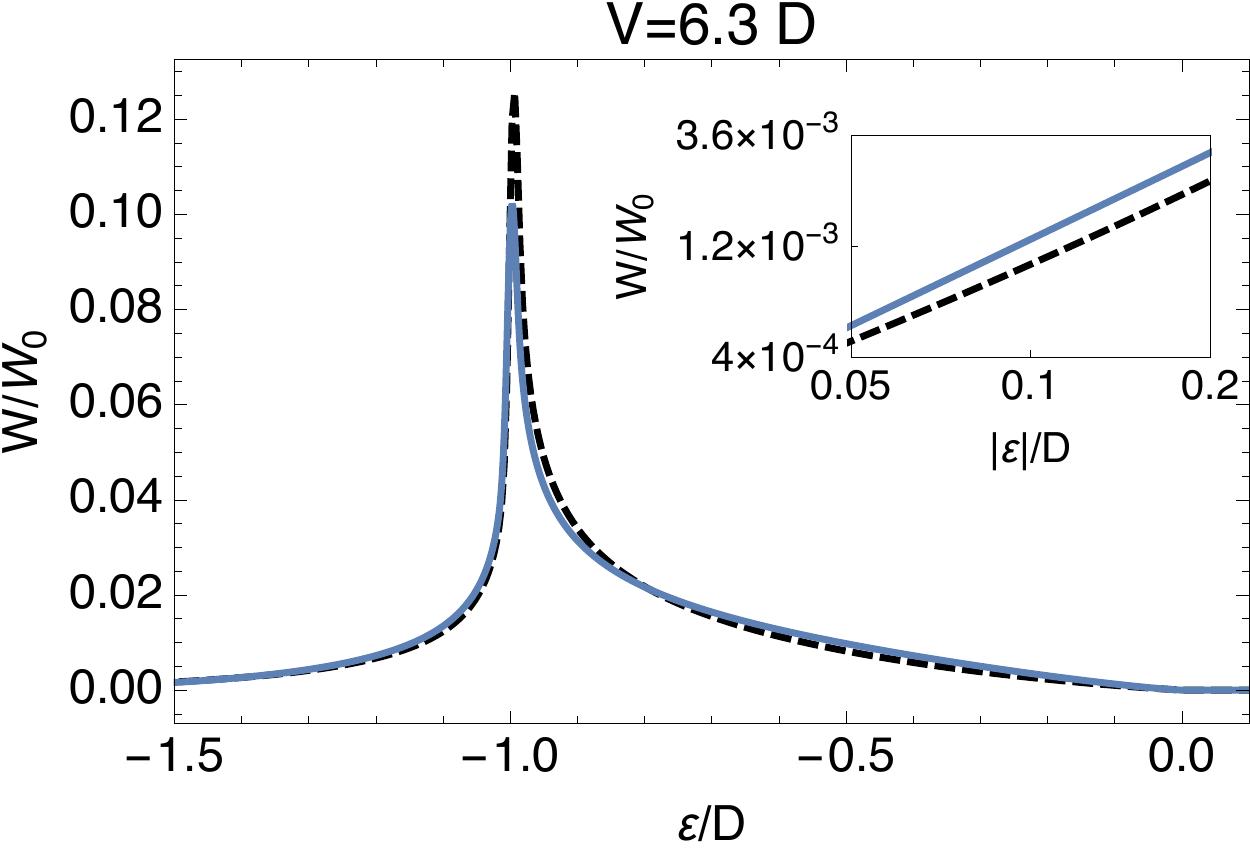}
\caption{X-ray transition rate $W(\ve)$, in units of $W_0=\gamma^2/2\pi D$, 
for a half-filled cosine band. The two panels correspond
respectively to the two large positive values of the impurity
interaction, $V=D$ and $V=6.3D$.
Solid blue lines represent the superposed Ansatz~(\ref{full_ansatz}) and 
dashed black lines represent numerically exact results.\label{f6}}
\end{center}
\end{figure*}
A final result, and a highlight of our study, is the transition rate $W(\ve)$ 
calculated at $V>0$ with the superposed Ansatz. As can be seen in Figure \ref{f6}, there is
excellent agreement with the exact result, up to $V$ as large as $6.3D$, which
corresponds to a phase shift $\phi(0)=0.9\, \pi/2$. Not only do we capture
the Fermi-edge singularity for $\ve\to0$, but multiple-particle excitations corresponding to
$\ve<-D$ are also well-accounted for.
This implies that our
variational treatment accurately describes all excitations that are produced when
the Fermi sea is shaken up during an inelastic tunnelling transition, even at
strong impurity interactions and large energy transfers.

\section{Conclusion and perspectives}

While bosonization has proven invaluable in the study of impurity and bulk one-dimensional
systems~\cite{Giam}, its practical applications has until now largely been confined 
to the study of linearly dispersing fermion models. More realistic calculations
in the fermionic language are typically based on an infinite size 
extension~\cite{Vidal,Moore} of the density matrix renormalization 
group~\cite{White,Schollwoeck}, leading for instance to detailed studies of quantum 
spin chains~\cite{Kollath}.
Our work provides a proof of principle that the bosonization technique can be successfully 
applied to lattice models, incorporating a non-trivial band structure into a description valid on all energy 
scales.

As a testing ground for our ideas, we 
studied the x-ray edge singularity for a microscopic model of a
one-dimensional electronic band, investigating the emission rate on all energy
scales. We found, in addition to the known Fermi-edge singularity, a strong reorganization
of many-electron states even at energy scales below the band edge, that can be accounted for
simply in a bosonic language. At weak coupling, we found an analytic expression for the
x-ray spectrum, that is quantitatively correct at all scales, and that, to our best knowledge,
has not appeared in the literature before. At strong coupling we provide a non-perturbative
variational solution, that quantitatively accounts for x-ray processes even at frequencies outside the
conduction band, where the signal is exclusively produced by multi-electron processes. 
Besides building intuition
by efficiently parameterizing Slater determinants in terms of bosonic coherent
states, our scheme turned out to be numerically very efficient, 
with a scaling $\Omega\ln\Omega$ of the
computation time, where $\Omega$ is the size of the energy grid used to discretize the system, 
instead of $\Omega^3$ for the exact diagonalization. 
Though not exact, the variational approach 
yields very accurate results (see our main result in Figure \ref{f6}).

On a conceptual level, this leads to a drastic change of viewpoint in the many-body problem, 
in which a technique usually associated with effective low-energy 
theories, is used to predict microscopically the behavior of excitations across 
the whole band structure for lattice electronic models with arbitrary band dispersion.
In future work we plan to exploit this to study bulk-interacting fermions on a lattice.
Another avenue for future work is to extend the coherent state variational technique
employed here to deal with time-dependent problems, as has already been done in
the context of the spin-boson model~\cite{Gheeraert}.

Our analysis was based on trial wave functions built on coherent states. In future 
applications of our bosonization approach, the many-boson systems that results could in 
principle be studied by other numerical means as well (quantum Monte
Carlo~\cite{Moon,Leung,Hamamoto,Rosch} 
or the Numerical Renormalization Group~\cite{Freyn}). An interesting open question
concerns whether the microscopic bosonic description also holds advantages over
the original fermionic one, as it did in the case we studied, when applied in conjunction 
with these techniques.  

\begin{appendices}
\section*{Appendix: Physical meaning of the coherent state wave function} 
\label{meaning}
We address here the meaning of the bosonic coherent states in terms of the 
original fermions for the problem of the dispersive band, in first quantization 
language. The bosonic coherent state $\left|f\right>$ (\ref{restricted}) is the exact
ground state of the bosonic parent Hamiltonian
\begin{equation}
H_f=\int_{0}^\infty d\ve\,\left[\ve b^\dagger_\ve b_\ve + \sqrt{\ve} f(\ve)(b_\ve+b_\ve^\dagger)\right].
\end{equation}
Using the relation (\ref{bdef}) between the bosonic $b_\ve$ operators and the
fermionic $\psi(\tau)$ operators, the parent Hamiltonian can be refermionized 
\begin{equation}
H_f=\int_{-\infty}^\infty d\tau\,\left[:\psi(\tau)^\dagger(-i\partial_\tau)\psi(\tau):
+\bar f(\tau)\psi^\dagger(\tau)\psi(\tau)\right],
\end{equation}
where $\bar f(\tau)=\int_{-\infty}^\infty d\omega\,e^{i\omega t}f(|\omega|)$.
The exact x-ray edge Hamiltonian can be expressed in the $\tau$ basis as:
\begin{eqnarray}
\label{Htimedomain}
\hspace{-2cm}
 H_{V,{\rm enl}} =  -i \int_{-\infty}^{+\infty} \!\!d\tau \, \psi^\dagger(\tau)
\partial_\tau \psi(\tau)
+ 4\pi V\int_{-\infty}^{\infty} \!\!d\tau_1
\int_{-\infty}^{\infty}\!\!d\tau_2\, \Delta(\tau_1)
\Delta(\tau_2)^*\psi^\dagger(\tau_1)\psi(\tau_2).
\end{eqnarray}
Thus we see that approximating the ground state of the true Hamiltonian
(\ref{Htimedomain}) with a coherent state, amounts to replacing the non-local in
$\tau$ scattering term, with an effective potential $\int_{-\infty}^\infty
d\tau\,\bar f(\tau)\psi^\dagger(\tau)\psi(\tau)$ that is local in $\tau$.
If we denote the single-particle eigenstate with energy $\ve$ of this
non-interacting parent Hamiltonian by $\left|\psi_\ve\right>$, then the
time-representation single-particle wave functions read
\begin{equation}
\left<\tau\right.\left|\psi_\ve\right>=\frac{1}{\sqrt{2\pi}}e^{i\ve \tau - i \bar F(\tau)},
\end{equation}
where we may choose
\begin{equation}
\bar F(\tau)=\frac{1}{2}\int_{-\tau}^\tau d\tau' \bar f(\tau')=\int_{-\infty}^\infty d\omega \frac{f(|\omega|)\sin(\omega\tau)}{\omega}.
\end{equation}
Let us return to the energy representation and denote the single-particle
eigenstates of the clean system as
$c_\ve^\dagger\left|0\right>=\left|\ve\right>$. In the energy representation
the single-particle wave functions of the parent Hamiltonian read
\begin{equation}
\left<\ve'\right|\left.\psi_\ve\right>=\frac{1}{\sqrt{2\pi}}\int_{-\infty}^\infty 
d\tau e^{-i\ve'\tau}\left<\tau\right|\left.\psi_\ve\right>=\left<\ve'\right|e^{-iF}\left|\ve\right>.\label{matel}
\end{equation} 
Here $e^{-iF}$ is an operator that acts on the single-particle Hilbert space
such that $\left|\psi_\ve\right>=e^{-iF}\left|\ve\right>$. The operator $F$ has
matrix elements
\begin{equation}
\left<\ve'\right|F\left|\ve\right>=\int_{-\infty}^\infty \frac{d\tau}{2\pi}e^{i(\ve'-\ve)\tau}\bar F(\tau)=i\mathcal P\frac{f(|\ve'-\ve|)}{\ve'-\ve}.
\end{equation}
A Slater determinant, constructed from the single-particle wave functions
$\left<\ve'\right|\left.\psi_\ve\right>$ with $\ve<\ve_F$, is fully equivalent
to the single bosonic coherent state~(\ref{restricted}). Up to moderate
interaction strength $V$, there is little mixing of physical and spectator
degrees of freedom in the coherent state, implying that
$\left<\ve'\right|\left.\psi_\ve\right>$ is close to a delta-function for
$\ve'>|D|$. In this regime a Slater determinant restricted to band orbitals
only, and excluding spectator orbitals, is thus nearly equivalent to a bosonic
coherent state. 

\end{appendices}

\ack This work is based on research supported in part by the National Research Foundation 
of South Africa (Grant Number 90657) and CNRS PICS contract ``FermiCats''.

\setcounter{figure}{0}
\setcounter{table}{0}
\setcounter{equation}{0}
\setcounter{section}{0}

\global\long\def\theequation{S\arabic{equation}}
\global\long\def\thefigure{S\arabic{figure}}

\vspace{1.0cm}
\begin{center}
{\bf \large Supplementary information for
``Microscopic bosonization of
band structures: X-ray processes beyond the Fermi edge''}
\vspace{0.5cm}\\
\end{center}

\section{Exact diagonalization method for the x-ray edge problem}

In the main text we compare results obtained via our microscopic bosonization method to results obtained via 
exact diagonalization. Here we explain how the exact results were obtained. 

In order to develop a method leading to numerically exact results, we consider 
a lattice with $2\Omega-1$ sites, and assume periodic boundary conditions. The discrete 
version of $H_V$ reads
\begin{equation}
H_V=\sum_{m=0}^{\Omega-1} \ve(q_m) \tilde c_m^\dagger \tilde 
c_m+V\tilde\psi_0^\dagger\tilde\psi_0,~~~q_m=\frac{\pi m}{\Omega-\frac{1}{2}},
\end{equation}
where
\begin{equation}
\tilde \psi_0=\frac{1}{\sqrt{\Omega-\frac{1}{2}}}\sum_{m=0}^{\Omega-1}\frac{\tilde c_m}{\sqrt{1+\delta_{m0}}},
\end{equation}
and $\tilde c_m^\dagger$ creates a fermion in the even band orbital with
energy $\ve(q_m)$. We use Wick's theorem, to express $P(t)$ in terms of single
particle matrix elements. For a band containing $\mathcal N_+$ particles in the
even-mode single particle orbitals, this gives
\begin{equation}
P(t)=e^{i E_0 t}{\rm det}[M(t)]\sum_{m,n=0}^{\mathcal N_+-1}\left<\left.\tilde \psi_0\right| m,V\right>
\left[M(t)^{-1}\right]_{m,n}\left<n,V\left|\tilde \psi_0\right.\right>.
\end{equation}
In this expression, $M(t)$ is an $\mathcal N_+\times \mathcal N_+$ matrix with entries
\begin{equation}
M(t)_{m,n}=\left<m,V\right|e^{-i h_0 t}\left|n,V\right>
\end{equation}
and $\{\left|m,V\right>|m=0,\,1,\,\ldots,\,\mathcal N_+-1\}$ are the the lowest
$\mathcal N_+$ single-particle orbitals of $H_V$, {\it i.e.} the ones that are
occupied when the conductor is in the initial state $\left|\Psi_0^V\right>$. The
operator 
\begin{equation}
h_0=\sum_{m=0}^{\Omega-1}\varepsilon_m \left|m\right>
\left<m\right|,~~~~\left|m\right>=\tilde c_m^\dagger\left|0\right>
\end{equation}
is the single-particle Hamiltonian corresponding to $H_0$. Finally 
$\left|\tilde \psi_0\right>=\tilde \psi_0^\dagger\left|0\right>$ is the single-particle state
with a fermion localized in the Wannier orbital centred on site zero of
the crystal.
The matrix $M(t)$, its inverse and determinant are then calculated numerically
for a large number of discrete times. From there, $P(t)$ is calculated, and
integrated numerically to obtain $W(\ve)$. The computation time for a single
evaluation of $P(t)$ scales like $\mathcal N_+^3$ as a function of the number of
particles $\mathcal N_+$.

\section{Details on the variational optimization}
\label{app:numdet}

For the variational calculations, we implemented the numerical minimization using
the quasi-Newton method L-BFGS-B. The energy interval
$\varepsilon\in(0,\infty)$ is truncated to $\varepsilon\in(0,8D)$, and then
discretized into a regular lattice of $2^{13}$ points. During the minimization
process, the norm of the gradient of the energy functional drops to $10^{-7}D$
within a minute's running time on a desktop computer, even for the more
complicated superposed Ansatz~(\ref{full_ansatz}). We experimented with 
different initial conditions.
For instance, we compared what happens when one uses the optimal $f(\ve)$ of the
first Ansatz as an initial condition for the minimization of the second Ansatz,
to what happens when one uses $f(\ve)=0$. We have also compared the case where
the boundary condition $g(0)=f(0)-1$ is explicitly imposed, to the case where
$f(0)$ and $g(0)$ are treated as independent, and initially $f(0)-g(0)\not=1$.
These choices affect how the minimum is approached, for instance whether the
norm of the gradient decreases smoothly or noisily, but we always find the same
minimum. This suggests that the found minimum is unique. 
We were able to perform the brute force numerics for a crystal of $2^{10}-1$
sites. This allows us to calculate the exact $W(\ve)$ at an energy resolution
$\Delta E \sim 2^{-8} D$. In principle, the variational data corresponds to a
larger system. Thus, we convolve the variational and brute force rates
with the same Gaussian of width $\sim 2^{-8} D$, to eliminate differences that
are due to the poorer resolution of the exact results.

\section{Minimization of the superposed Ansatz}

In the main text, we expressed the energy functional and its gradient in terms
of the variational parameters of the single coherent state
Ansatz. (See (45) -- (49) of the main text.) Here we do the same for the superposed Ansatz~[(54) in main text]. 
From the outset, we assume 
a particle-hole symmetric dispersion relation so that $N(-\ve)=N(\ve)$ and $\Delta(t)$ 
is real. We place the Fermi energy at $\ve_F=0$, in the centre of the band.
For $V>0$, the energy functional to minimize is given by the formal
expression:
\begin{eqnarray}
\fl E_{\rm var}&=\frac{\left<\Psi_{\rm var}\right|H_{V,{\rm enl}}\left|\Psi_{\rm var}\right>}
{\left<\Psi_{\rm var}\right|\left.\Psi_{\rm var}\right>}\nonumber\\
\fl &=\frac{\left<f\right| H_{V,{\rm enl}}+V/2\left| f\right>+2c\left<f\right|H_{0,{\rm enl}}U^\dagger\tilde 
\psi_0\left|g\right>+c^2\left<g\right|\tilde \psi_0^\dagger H_{0,{\rm enl}} \tilde\psi_0\left|g\right>}
{1+2c\left<f\right|U^\dagger\tilde \psi_0\left|g\right>+c^2\left<g\right|\tilde 
\psi_0^\dagger\tilde \psi_0\left|g\right>}-\frac{V}{2}.
\end{eqnarray}
All the overlaps in the above expression are real. In the second and third
terms of the denominator in the right-hand-side term, we replaced 
Hamiltonian $H_{V,{\rm enl}}+V/2$ with $H_{0,{\rm enl}}$. This is allowed, because 
the term $V\tilde\psi_0^\dagger\tilde\psi_0$ in $H_{V,{\rm enl}}$ produces zero when
acting on the term $U^\dagger \tilde\psi_0\left|g\right>$ in the full Ansatz.
For the purpose of expressing $E_{\rm var}$ in terms of the variational
parameters $f(\varepsilon)$, $g(\varepsilon)$ and $c$, it is convenient to
use auxiliary functions $A(\omega)$, $B(\omega)$ and $C(\omega)$ as defined in (46), (60) and (61) in the main text.
In terms of the auxiliary functions, the overlaps appearing in the energy functional can 
finally be expressed as
\begin{equation}
\left<f\right|H_{V,{\rm enl}}+V/2 \left|f\right>=\int_0^\infty \!\!d\varepsilon\, 
f(\varepsilon)^2+2V\int_0^\infty \!\!d\omega \, A(\omega)^2, \label{e1}
\end{equation}
\begin{equation}
\left<f\right|U^\dagger\tilde \psi_0\left|g\right>=B(0),\label{e2}
\end{equation}
\begin{equation}
\left<f\right|H_{0,{\rm enl}}U^\dagger \tilde \psi_0\left|g\right>=B(0)\int_0^\infty 
\!\!d\varepsilon\,f(\varepsilon)g(\varepsilon)+\int_0^\infty \!\!
d\varepsilon\,f(\varepsilon)B(\varepsilon),\label{e3}
\end{equation}
\begin{equation}
\left<g\right|\tilde \psi_0^\dagger \tilde \psi_0\left|g\right>
=2\int_0^\infty \!\! d\omega\,C(\omega)^2,\label{e4}
\end{equation}

\begin{eqnarray}
\left<g\right|\tilde \psi_0^\dagger H_{0,{\rm enl}}\tilde \psi_0\left|g\right>
= 2\left(\int_0^\infty \!\!d\varepsilon\,g(\varepsilon)^2\right)\int_0^\infty 
\!\!d\omega\,C(\omega)^2\nonumber\\
+4\int_0^\infty \!\! d\varepsilon \int_0^\infty \!\!d\omega\,g(\varepsilon)
C(\omega)C(\omega+\varepsilon)+2\int_0^\infty \!\! d\omega\,\omega C(\omega)^2.
\label{e5}
\end{eqnarray}
To use a quasi-Newton minimization scheme, we also need to know the gradient of
the energy functional with respect to the variational parameters. The expression
for $\partial E_{\rm var}/\partial c$ is straight-forward and we do not write it
out explicitly. The functional derivatives of $E_{\rm var}$ with respect to
$f(\varepsilon)$ and $g(\varepsilon)$ can be constructed from the following
parts, together with (\ref{e1})--(\ref{e5}):
\begin{eqnarray}
\fl &\frac{\delta}{\delta f(\varepsilon)}\left<f\right|H_{V,{\rm enl}}+V/2\left|f\right>
=2f(\varepsilon)
+\frac{4V}{\varepsilon}
\int_0^\infty \!\! d\omega\,
A(\omega)\left[A(\omega+\varepsilon)-A(\omega-\varepsilon)\right],
\end{eqnarray}
\begin{equation}
\fl \frac{\delta}{\delta f(\varepsilon)}\left<f\right|U^\dagger\tilde \psi_0\left|g\right>
=\frac{g(\varepsilon)-f(\varepsilon)}{\varepsilon}B(0)+\frac{B(\varepsilon)}{\varepsilon},
\end{equation}
\begin{equation}
\fl \frac{\delta}{\delta g(\varepsilon)}\left<f\right|U^\dagger\tilde \psi_0\left|g\right>
=\frac{f(\varepsilon)-g(\varepsilon)}{\varepsilon}B(0)-\frac{B(-\varepsilon)}{\varepsilon},
\end{equation}
\begin{eqnarray}
\fl &\frac{\delta}{\delta f(\varepsilon)}\left<f\right|H_{0,{\rm enl}}U^\dagger\tilde \psi_0\left|g\right>
=g(\varepsilon)B(0)
+\left[\frac{g(\varepsilon)-f(\varepsilon)}{\varepsilon}B(0)+\frac{B(\varepsilon)}{\varepsilon}\right]
\int_0^\infty \!\! d\omega\,f(\omega)g(\omega)\nonumber\\
\fl &~~~~+B(\varepsilon)
+\frac{g(\varepsilon)-f(\varepsilon)}{\varepsilon}\int_0^\infty \!\! d\omega\,f(\omega)B(\omega)
+\frac{1}{\varepsilon}\int_0^\infty \!\!
d\omega\,f(\omega)B(\omega+\varepsilon),
\end{eqnarray}
\begin{eqnarray}
\fl &\frac{\delta}{\delta g(\varepsilon)}\left<f\right|H_{0,{\rm enl}}U^\dagger 
\tilde\psi_0\left|g\right>=f(\varepsilon)B(0)
+\left[
\frac{f(\varepsilon)-g(\varepsilon)}{\varepsilon}B(0)-\frac{B(-\varepsilon)}{\varepsilon}\right]
\int_0^\infty \!\! d\omega\,f(\omega)g(\omega)\nonumber\\
\fl &~~~~+\frac{f(\varepsilon)-g(\varepsilon)}{\varepsilon}\int_0^\infty \!\! d\omega\,f(\omega)B(\omega)
-\frac{1}{\varepsilon}\int_0^\infty \!\!
d\omega\,f(\omega)B(\omega-\varepsilon),
\end{eqnarray}
\begin{equation}
\fl \frac{\delta}{\delta g(\varepsilon)}\left<g\right|\tilde \psi_0^\dagger \tilde \psi_0\left|g\right>
=\frac{4}{\varepsilon} \int_0^\infty \!\! d\omega\,
C(\omega)\left[C(\omega+\varepsilon)-C(\omega-\varepsilon)\right],
\end{equation}
\begin{eqnarray}
\fl &\frac{\delta}{\delta g(\varepsilon)}\left<g\right|\tilde \psi_0^\dagger H_{0,{\rm enl}}\tilde 
\psi_0\left|g\right> =4 g(\varepsilon)\int_0^\infty \!\! d\omega\,C(\omega)^2\nonumber\\
\fl &~~~~~+\frac{4}{\varepsilon}\left(\int_0^\infty \!\!
d\omega\,g(\omega)^2\right)\int_0^\infty \!\! d\omega\,
C(\omega)\left[C(\omega+\varepsilon)-C(\omega-\varepsilon)\right]+4\int_0^\infty \!\! d\omega\,C(\omega)C(\omega+\varepsilon)\nonumber\\
\fl &~~~~~+\frac{4}{\varepsilon}\int_0^\infty \!\! d\omega \int_0^\infty d\bar{\omega}\,
g(\bar{\omega})\left[C(\omega+\varepsilon)-C(\omega-\varepsilon)\right]C(\omega+\bar\omega)\nonumber\\
\fl &~~~~~+\frac{4}{\varepsilon}\int_0^\infty \!\!d\omega \int_0^\infty \!\! d\bar{\omega}\,
g(\bar{\omega})C(\omega)\left[C(\omega+\bar{\omega}+\varepsilon)-C(\omega+\bar{\omega}-\varepsilon)\right]\nonumber\\
\fl &~~~~~+\frac{4}{\varepsilon}\int_0^\infty \!\! d\omega\,\omega\,C(\omega)
\left[C(\omega+\varepsilon)-C(\omega-\varepsilon)\right].
\end{eqnarray}
To evaluate these expressions numerically, we converted convolution-type
integrals into sequences of Fourier type transforms. For instance
\begin{eqnarray}
\int_0^\infty \!\! d\omega &\int_0^\infty \!\! d\bar{\omega}\,g(\bar{\omega})
\left[C(\omega+\varepsilon)-C(\omega-\varepsilon)\right]C(\omega+\bar\omega)\nonumber\\
&=-2i\int_{-\infty}^\infty \!\! d\tau\, \sin(\varepsilon\tau) Q(\tau)R(\tau),\label{mult}
\end{eqnarray}
with the auxiliary functions:
\begin{eqnarray}
Q(\tau)&=&\int_{-\infty}^\infty
\!\!\frac{d\omega}{2\pi}\,e^{i\omega\tau}C(\omega),~~
R(\tau)=\int_0^\infty \!\!d\omega\, e^{-i\omega\tau} S(\omega),\\
S(\omega)&=&\int_{-\infty}^\infty \!\!d\tau\,e^{-i\omega\tau}Q(\tau)T(\tau),~~
T(\tau)=\int_0^\infty \!\!d\omega\,e^{-i\omega\tau} g(\omega).
\end{eqnarray}

When one uses this strategy, together with a fast Fourier algorithm, the
execution time for one evaluation of the energy and its gradient scales like
$\Omega{\rm ln}(\Omega)$, where $\Omega$ is the number of discretized modes
$\varepsilon_n$. In contrast, a naive evaluation of (\ref{mult}) for every mode
$\varepsilon_n$ would have an execution time that scales like $\Omega^3$.

\end{document}